\documentclass[draft]{agujournal}
\draftfalse
\journalname{JGR-Space Physics}
\usepackage{amsmath}
\usepackage{amsfonts}
\usepackage{amssymb}
\usepackage{graphicx}
\usepackage{verbatim}%

\begin{document}

\title{Formation of plasma around a small meteoroid: 1. Kinetic theory}

\authors{Y.~S.~Dimant\affil{1}\  and M.~M.~Oppenheim\affil{1}}
\affiliation{1}{Center for Space Physics, Boston University}

\correspondingauthor{Y.~S. Dimant}{dimant@bu.edu}

\begin{keypoints}
\item Develops the first kinetic theory of plasma formed around a
  small ablating meteoroid
\item Obtains analytical expressions for the spatial and
  velocity distributions of ablated ions and neutrals
\item Provides a basis for quantitative interpretation of radar head-echo measurements
\end{keypoints}

\begin{abstract}
Every second millions of small meteoroids enter the Earth's  atmosphere
producing dense plasmas. Radars easily detect these  plasmas and researchers
use this data to characterize both the  meteoroids and the atmosphere. This
paper develops a  first-principle kinetic theory describing the behavior of
particles,  ablated from a fast-moving meteoroid, that colliside with the
atmospheric molecules. This theory produces analytic expressions describing
the spatial structure and velocity distributions of ions and neutrals near the ablating meteoroid.
This analytical model will serve as a  basis for a more accurate quantitative
interpretation of radar  measurements and should help calculate meteoroid and
atmosphere  parameters from radar head-echo observations.

\end{abstract}

\section{Introduction}

\begin{figure}[h]
\centering
\includegraphics[width=30pc]
{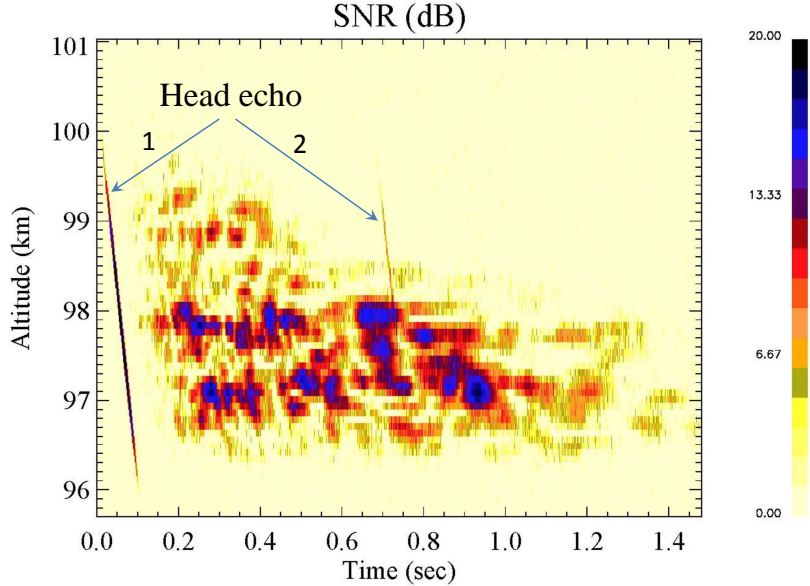}
\caption{Altitude-time radar image of meteor plasma (JRO, July 12, 2005,
3:43AM LT). The color coding shows the signal-to-noise ratio in dB. The two
arrows point to the head echoes of two separate meteoroids. The stronger head
echo (1) is followed by a non-specular trail, while the meteoroid with the
much weaker head echo (2) produces no detectable trail. The vertical velocity
components of both meteoroids were close. The head-echo slope gives the
corresponding downward speed about $40$~km/s.}
\label{Fig:HeadEcho}
\end{figure}

Every second millions of tiny, submilligram and submillimeter, meteoroids hit
the Earth, depositing tons of extra-terrestrial material in its atmosphere.
The majority of these particles do not create visual signatures but large
radars, such as at Arecibo and Jicamarca, can often detect many particles per
second despite only scanning a few square kilometers. These radars do not
measure the meteoroids themselves but instead detect the plasma generated as
they ablate, making measurements called head echoes. Figure~\ref{Fig:HeadEcho}
shows an example of one such measurement. Interpreting these measurements
requires a quantitative understanding of the structure of the neutral gas and
plasma surrounding a meteoroid. This paper develops a first-principle kinetic
model aimed at interpreting meteor head echo signals
\citep{Bronshten:Physics83,Ceplecha:Meteor1998,Close:NewMethod2005,Campbell-Brown:Meteoroid2007}.

Determining the composition of small meteoroids has proven difficult. By
analogy with bigger meteorites that reach the Earth's surface. researchers
assume that small meteoroids are composed of free metals like iron, nickel,
cobalt, volatiles like carbon, water, sulphur, and mineral oxides like FeO,
SiO$_{2}$, MgO, etc. Optical spectral measurements of meteors corroborate this
assumption but cannot say much about elements that do not have strong spectral
signatures \citep{Borovicka:Fireball93}. A variety of techniques have led researchers to estimate
that the meteoroid mass distribution peaks at around $1~\mu$g
\citep{Grun:Collisional85,Close:Meteor07,Blaauw:Mass11,
Fucetola:Meteoroid16}.

Meteoroids reach the Earth at hypersonic speeds, $U=11$--$73$ km/s, and become
detectable when they encounter sufficiently dense air to heat them up through
friction. This results in some sputtering but primarily sublimation from the
surface in a process called ablation. This forms a mostly neutral gas cloud
around the meteoroid. High-velocity collisions with atmospheric molecules
partially ionize and decelerate this gas, forming a dense meteor plasma.

Typical small meteoroids become visible to radars at $h\simeq120$~km altitude
and disappear below $75$ km \citep{Janches:Initial2005}. At these altitudes,
the fast-descending meteoroid leaves behind a plasma column that lives for a
relatively long time until it diffuses, disintegrates and, eventually,
recombines. The lowest altitude where meteors typically disappear from radar
observations roughly marks the altitude where meteoroids have either
disintegrated or decelerated to the point where they stop generating plasma.

Meteor plasma is usually a few orders of magnitude denser than the ambient
ionosphere, especially during a night time.  High-power large aperture (HPLA)
radars located near the magnetic equator, such as Jicamarca Radio Observatory
(JRO) or ALTAIR, often detect signals composed of two distinct parts, the head
echo and non-specular (range-spread) trail, as shown in
Figure~\ref{Fig:HeadEcho}. On an altitude-time diagram, the head echo
resembles an almost straight line. It forms when a radar beam scatters from a
dense plasma that accompanies an ablating meteoroid. The reflected signal has
high Doppler shift, showing that the plasma moves at or near the velocity of
the meteoroid.

This paper analyzes the structure of short-lived near-meteoroid plasmas that
produce radar head echoes. An accurate model of this structure will enable us
to better use these measurements to estimate meteoroid characteristics
\citep{Bronshten:Physics83,Ceplecha:Meteor1998}. This work has also
another motivation. Big meteor fireballs produce strong electromagnetic pulses
that result in audible sounds called electrophonics \citep{Bronshten:Physics83,Ceplecha:Meteor1998,
Keay:Continued95,Zgrablic:Instrumental02,Chakrabarti:VLF05,Lashkari:Detecting15}.
Recent ground-based antenna observations during
meteor storms have demonstrated that small, optically invisible, meteoroids
can also produce detectable electromagnetic pulses \citep{Price:ELF00,Rault:Searching10,
De:Detection11,Guha:Investigation12,Obenberger:Detection14}. To understand the physical nature of these
pulses and some of their non-trivial spectral features, it is desirable to
quantitatively understand the transient electric current system associated
with the near-meteoroid plasma. The kinetic theory of this paper should help
model this system. It could also predict whether the near-meteoroid plasma can
develop plasma instabilities that might be responsible for submillisecond
modulations of the observed ELF/VLF pulses \citep{Price:ELF00}.

Many previous theoretical studies of meteor plasma were devoted to estimating
the parameter $\beta$ that characterizes the meteor ionization efficiency, the
number of electrons generated by a meteoroid of a given speed and altitude per
unit length \citep{Massey:Collision55,Furman:The_theory_I_1960,Furman:A_theory_II_1961,
Furman:Notes_III_1964,Furman:Meteor-Trail_IV_1967,Lazarus:Meteor63,Sida:Production69,
Jones:Theoretical97,Bronshten:Physics83}; see also \citet{Ceplecha:Meteor1998} and references therein.
One of the earliest works that dealt with the spatial structure of the ablated
meteoric material is that of \citet{Manning:Initial58} who modelled the
initial radius of the collisionless kinetic expansion of a cylindrical trail
followed by its regular collisional diffusion. This model was based on
simplified assumptions like elastic sphere collisions, random walks leading to
Gaussian trails, meteoric ions ablated directly from the meteoroid, etc. The
most cited work by \citet{Jones:Theory1995} re-examined and developed
this approach further using both theory and particle simulations. Jones's
simulations have demonstrated significant deviations of the initial plasma
from the Gaussian profile across the plasma trail. The major goal of those
earlier studies was to model the 2-D transition from the initial ballistic
expansion to the regular collisional diffusion of the cylindrical plasma trail
behind the meteoroid. These studies were based on probabilistic averaging and
they did not really study the 3-D spatial distribution of the dense plasma
that surrounds the fast-moving meteoroid and is responsible for the radar head echo.

The meteor plasma that causes the head echo is distributed within a small
near-meteoroid volume whose effective size is of order of the collisional mean
free path. This and some other factors lead to strongly non-Maxwellian
velocity distributions and require a kinetic description. However, none of the
earlier studies have developed a consistent kinetic theory. Recently,
\citet{Dyrud:Plasma2008EMaP,Dyrud:Plasma2008AdSpR} made a first
attempt of modeling the spatial structure of the near-meteoroid plasma
kinetically using particle-in-cell (PIC) simulations. Their modeling generated
a plausible qualitative picture of what might be expected but did not provide
quantitative characteristics or simple parameter dependences that would allow
radar signal modeling and related physical analysis. The more recent paper by
\citet{Stokan:Particle-15} modeled collisions of ablated particles but, similar
to the early studies, it mostly studied the wake formation behind the
meteoroid. Our earlier theoretical papers mostly analyzed the slowly
diffusing plasma columns related to the non-specular trails
\citep{DimantOpp:MeteorTrail:1_Simul2006,DimantOpp:MeteorTrail:2_Theory2006}.

This paper develops a kinetic theory of the plasma formed around a
fast-descending meteoroid. Based on this theory, in a companion paper, we
calculate the 3-D spatial structure of the near-meteoroid plasma density
required for interpretation of radar head echoes. Future publications will
include a more detailed theoretical analysis, computer simulations, and
discussion of implications for real meteors and comparisons with radar observations.

This paper is organized as follows. In
section~\ref{Physical conditions and principal theoretical approximations} we
discuss the physical conditions of the actual meteors in atmosphere and our
choice of the relevant theoretical assumptions. In
section~\ref{Formulation of the kinetic problem} we formulate the kinetic
equation with the appropriate collisional operator, while in
section~\ref{Schematics of the general solution} we describe the general
approach to its solution. In
section~\ref{Velocity distribution of primary particles} we calculate the
distribution function of the primary, i.e., ablated meteoric particles that
have not yet experienced any collisions with the dense atmosphere. In
section~\ref{Velocity distribution of secondary particles} we calculate the
distribution function of the secondary particles (newly born ions and
singly-scattered neutrals). This section describes the kinetics of almost all
near-meteoroid ions and is central to the paper. In
section~\ref{Summary and conclusions} we summarize our objectives and findings.
Appendix~provides some mathematical details related to
section~\ref{Velocity distribution of secondary particles}.

\section{Physical conditions and principal theoretical
assumptions\label{Physical conditions and principal theoretical approximations}%
}

In this section, we analyze physical conditions of the meteor plasma formation
and justify our theoretical approximations. Readers who are not interested in
this justification can proceed directly to
section~\ref{Formulation of the kinetic problem}.

A fast descending meteoroid creates around itself a strong disturbance which
we will call the `near-meteoroid sheath.' It includes several components: (1)
the `primary,' mostly neutral, particles ablating from the meteoroid surface
and continuing their free ballistic motion until they collide with atmospheric
molecules; (2) collisionally scattered and ionized particles of both meteoric
and atmospheric materials; (3) free electrons released during the ionizing
collisions. This paper develops the kinetics of the heavy particles of
components (1) and (2).

Our analytical approach assumes that the kinetic energy of the ablated atoms
in the rest frame of the atmosphere greatly exceeds the thermal energy of the
ablated atoms and this greatly exceeds the thermal energy of the neutral
atmosphere. Hence,
\begin{linenomath*}
\begin{equation}
T_{\mathrm{A}}\ll T_{\mathrm{M}}\ll\frac{m_{\mathrm{A,M}}U^{2}}{2},
\label{hierarchy_energies}%
\end{equation}
\end{linenomath*}
where $T_{\mathrm{A}}$ is the undisturbed atmospheric temperature,
$T_{\mathrm{M}}<1\,\mathrm{eV}$ is the characteristic temperature of the ablated meteoric
particles (both in the energy units), $m_{\mathrm{A,M}}\sim(30$--$50)\,\mathrm{amu}$ are the corresponding
molecular or atomic masses, and $U=|\vec{U}|$ is the meteoroid speed. It also
assumes a tiny meteoroid, much smaller than any other scales in the
system, such that,
\begin{linenomath*}
\begin{equation}
r_{\mathrm{M}}\ll\lambda\ll H,L. \label{hierarchy_of_scales}%
\end{equation}
\end{linenomath*}
where $r_{\mathrm{M}}$ is the average meteoroid radius, $\lambda$ is the
characteristic length scale of the near-meteoroid sheath (of order of the
collisional mean free path), and $H$ is the scale height of atmosphere, and
$L$ is a characteristic length scale of spatial variations along the meteoroid
trajectory such as the meteoroid velocity, $-\vec{U}$, temperature, and
material composition.

The inequality $r_{\mathrm{M}}\ll\lambda$ precludes small meteoroids from
forming shock waves. The inequalities $\lambda\ll H,L$, suggest a simple
adiabatic approach when over the characteristic length of the near-meteoroid
sheath $\sim\lambda$ at a given altitude $h$, we can treat the atmosphere as a
uniform gas with the local density $n_{\mathrm{A}}$ and constant meteoroid
parameters.
Differential ablation \citep{von_Zahn:Iron1999,Vondrak:Chemical2008,Janches:First2009}
and meteoric fragmentation \citep{Kero:Three2008,Roy:Genetic2009,Mathews:Extensive2010} might in principle break
this gradual adiabatic behavior. However, these processes present no real
challenge to our theory.

Additionally, we will assume that the sheath particles collide almost
exclusively with the undisturbed neutral atmosphere of a given density and
composition, rather than between themselves. For simplicity, we will also
assume that the undisturbed neutral atmosphere includes only one atmospheric
species with the known density, $n_{\mathrm{A}}(h)$, mean molecular mass,
$m_{\mathrm{A}}$, and given velocity-dependent collision frequency. We will
also neglect inelastic collisions with the energy losses, $\Delta
E_{\mathrm{in}}\lesssim10$~eV, compared to
\begin{linenomath*}
\begin{equation}
\frac{m_{\mathrm{A,M}}U^{2}}{2}\approx140\,\mathrm{eV}\left(\frac
{m_{\mathrm{A,M}}}{30\,\mathrm{amu}}\right)  \left(  \frac{U}{30\,\mathrm{km}
/\mathrm{s}}\right)^{2}.
\label{huge_energy}%
\end{equation}
\end{linenomath*}	
We will initially include $\Delta E_{\mathrm{in}}$ for estimates and future
references, but in the rest of our theory we will neglect it.

Our most serious assumption is that between consecutive collisions ions are
weakly affected by fields and have almost straight-line ballistic trajectories
as do neutral particles. In this paper, we neglect completely the electric and
magnetic fields on the ion orbits. This can be justified by considering the
electron dynamics. Electrons released during the collisional ionization of
neutral particles form the vast majority of free electrons around the
descending meteoroid. Their characteristic kinetic energy, $\mathcal{E}_{e}$,
is typically below a few electronvolts with the corresponding electron speed
given by
\begin{linenomath*}
\[
V_{e}=\sqrt{\frac{2\mathcal{E}_{e}}{m_{e}}}\approx1.3\times10^{3}%
\ \mathrm{km}/\mathrm{s}\left(  \frac{\mathcal{E}_{e}}{5\,\mathrm{eV}}\right)
^{1/2},
\]
\end{linenomath*}
where $m_{e}$ is the electron mass. This speed far exceeds the hypersonic
meteoroid speed of $U\lesssim70~\mathrm{km/s}$, so that after release from a
neutral particle, the electron could travel long distances away from the
meteoroid unless it is confined by the geomagnetic field, $\vec{B}$, and an
electrostatic field, $\vec{E}=-\nabla\Phi$. At altitudes $\gtrsim
80\ \mathrm{km}$ electrons are highly magnetized with the characteristic
Larmor radius given by%
\begin{linenomath*}
\[
\rho_{e}\approx 0.3\,\mathrm{m}\left(\frac{\mathcal{E}_{e}}{5\,\mathrm{eV}%
}\right)  ^{1/2}\left(  \frac{2.5\times10^{-5}\,\mathrm{T}}{B}\right)  ,
\]
\end{linenomath*}
where $2.5\times10^{-5}\,\mathrm{T}$ is the typical value of the magnetic
intensity, $B=|\vec{B}|$, at the geomagnetic equator. Such relatively small
values of $\rho_{e}$ force the newly born electrons to remain practically
attached to the original magnetic field line (in the Earth's frame), but these
electrons still could travel freely along $\vec{B}$. However, as electrons
move away from the much slower ions they immediately create a charge
separation that builds up for them a retarding electrostatic barrier. The
corresponding ambipolar electric field is the main force that distorts the
orbits of free ions.

We can estimate this potential barrier by assuming a Maxwellian electron
velocity distribution with the uniform effective temperature, $T_{e}$, in the
few-electronvolts range (e.g., $T_{e}\sim3\ \mathrm{eV}$). This velocity
distribution translates to the Boltzmann distribution of the electron density,
$n_{e}(\vec{r}_{1})/n_{e}(\vec{r}_{2})\simeq n_{0}\exp(e\Delta\Phi/T_{e})$,
where $n_{e}(\vec{r}_{1,2})$ are the electron densities in two different
locations and $\Delta\Phi=\Phi(\vec{r}_{1})-\Phi(\vec{r}_{2})$ is the
corresponding potential difference. The built-up ambipolar electrostatic
field, $\vec{E}=-\nabla\Phi$, retards electrons but accelerates ions. Assuming
the quasi-neutrality, $n_{e}\simeq n_{i}$, we can easily estimate the
potential difference between a given location inside the dense electron sheath
and its edge. Loosely defining this edge by setting the meteor-plasma density
to be of order of the background ionospheric density, $n_{\mathrm{ionos}}$, to
a logarithmic accuracy we obtain%
\begin{linenomath*}
\begin{equation}
e\Delta\Phi\simeq T_{e}\ln\frac{n_{i}}{n_{\mathrm{ionos}}}. \label{Delta_Phi}%
\end{equation}
\end{linenomath*}
Assuming the meteor plasma to be a few orders of magnitude denser than the
background ionosphere, we obtain that $e\Delta\Phi\lesssim30~$\textrm{eV}.
Since this $e\Delta\Phi$ is noticeably smaller than the characteristic ion
kinetic energy $\mathcal{E}_{i}\sim m_{i}U^{2}/2$ given by
equation~(\ref{huge_energy}) then to a zero-order approximation we can neglect
the $\vec{E}$-effect on ions.

There is an additional field that can affect the ion trajectories. If the
descending meteoroid with the velocity $-\vec{U}$ crosses the magnetic field
$\vec{B}$ then in its frame $\vec{B}$ induces the dynamo electric field
$\vec{E}=\vec{U}\times\vec{B}$. At the altitudes of $80\ \mathrm{km}\lesssim
h<120\ $km, ions are almost unmagnetized due to frequent collisions with the
atmosphere, $\Omega_{i}\ll\nu_{i}$, where $\Omega_{i}\ $is the ion
gyrofrequency. The characteristic size of the near-meteoroid plasma sheath,
$\lambda$, is of order of the ion mean free path, $\lambda\lesssim U/\nu_{i}$.
Then the corresponding potential difference across the sheath is $e\Delta
\Phi\sim eE\lambda\lesssim\left(  \Omega_{i}/\nu_{i}\right)  m_{i}U^{2}%
\ll\mathcal{E}_{i}$. Thus the induced electric field is also relatively weak.

In this paper we neglect $\vec{E}$ completely and assume the sheath electrons
to roughly follow the Boltzmann distribution with quasi-neutrality. This
closes the kinetic description of ions and allows ignoring the geomagnetic
field, $\vec{B}$. This in turn leads to the axial symmetry around the
straight-line meteoroid trajectory. We will improve our theoretical model in
future by using computer simulations.

To conclude this section, we note that most ions should originate from the
ablated meteoric particles due to their lower ionization potential, but a
certain fraction of the atmosphere can also be collisionally ionized.
Additionally, particles ablating from the meteoroid surface should be mostly
neutral, but a small fraction of them could be ionized. The general kinetic
framework of this paper encompasses all these possibilities.

\section{Formulation of the kinetic
problem\label{Formulation of the kinetic problem}}

The description of the near-meteoroid sheath around a descending meteoroid
requires a kinetic theory for two main reasons: (1) the characteristic length
scales of the near-meteoroid sheath are of order of the collisional mean free
paths and (2) the two colliding velocity distributions -- the ablative flow
from the meteoroid surface and the impinging neutral atmosphere -- are shifted
with respect to each other by a huge hypersonic velocity, $\vec{U}$. According
to (1), the near-meteoroid sheath forms a marginally collisional structure. It
includes mostly the `primary' (ablated) particles that travel freely before
suffering their first collision and the `secondary' particles, i.e., particles
scattered or ionized only once since the original ablation. According to (2),
after one or two collisions the two impinging flows redistribute their huge
energy difference only partially, so that the velocity distributions of the
sheath particles become neither isotropic nor Maxwellian. No fluid model can
adequately describe such distributions.

The primary interest of this paper lies in the near-meteoroid plasma, but
because we neglect field effects, as described in
section~\ref{Physical conditions and principal theoretical approximations}, the
kinetic framework of this paper does not distinguish between the
near-meteoroid plasma and the neutral sheath. Between consecutive collisions,
all heavy meteoric particles move over straight-line ballistic trajectories,
although the ion-neutral collisions differ from the neutral-neutral
collisions. We presume given expressions for each collisional cross-section.
Apart from these specific expressions, the sheath ions and neutral particles
are described by the same general equations.

Now we consider a group of the sheath particles with the given material and
charge state. This group is characterized by the velocity distribution
$f(\vec{V},\vec{r})$, so that the corresponding particle and flux densities
are given by $n=\int f(\vec{V},\vec{r})d^{3}V$ and $\vec{\Gamma}=\int
f(\vec{V},\vec{r})\vec{V}d^{3}V$, respectively. We do not index variables with
the explicit group identifiers because the general equations written below are
applicable to any group. We will develop our kinetic theory based on the
collisional kinetic equation derived in
Appendix~A. This equation generalizes the
standard kinetic equation with the Boltzmann collision operator
\citep{Huang:Statistical87,Lifshitz:Physical1981} to inelastic
collisions, albeit in a simplified form, as described below. Under stationary
conditions in the meteoroid frame of reference, the initial kinetic
equation~with binary collisions can be written as
\begin{linenomath*}
\begin{equation}
(\vec{V}\cdot\nabla)f=-\nu(\vec{V})f+\hat{S}_{\mathrm{gain}}[f^{\prime}].
\label{D_t(f)}%
\end{equation}
\end{linenomath*}
The left-hand side (LHS) of equation~(\ref{D_t(f)}) describes\ the ballistic
motion of the particles between the consecutive collisions.

The right-hand side (RHS) of equation~(\ref{D_t(f)})\ describes the binary
collision operator. The first term describes the collisional \textquotedblleft
loss\textquotedblright\ from a given elementary volume of the 6D phase space
\citep{Lifshitz:Physical1981}, $d^{3}Vd^{3}V_{\beta}$, where $\vec{V}$ is the
3D vector of the meteoric particles described by the given distribution
function $f$, while $\vec{V}_{\beta}$ is the velocity of the colliding
partners (atmospheric particles) described by $f_{\beta}$, and $\nu(\vec{V})$
is the velocity-dependent kinetic collision frequency,%
\begin{linenomath*}
\begin{equation}
\nu(\vec{V})=\sum_{\beta}\int f_{\beta}G\left(  u,\Lambda\right)
ud^{3}V_{\beta}d\Omega_{\mathrm{s}}\approx2\pi n_{\mathrm{A}}\int_{-1}%
^{1}uG\left(  u,\Lambda\right)  d\Lambda, \label{nu(V)}%
\end{equation}
\end{linenomath*}
where $u=|\vec{u}|$ is the relative speed of the two colliding particles and
$\Lambda$ is the cosine of the corresponding scatter angle, as will be
explained below. The second term in the RHS of equation~(\ref{D_t(f)}) is the
\textquotedblleft gain\textquotedblright\ component of the collision operator.
It is written as an integral operator $\hat{S}_{\mathrm{gain}}$ acting on
$f^{\prime}$, and expresses the collisional kinetic arrival at $d^{3}%
Vd^{3}V_{\beta}$ from other elementary volumes, $d^{3}V^{\prime}d^{3}V_{\beta
}^{\prime}$.
\begin{linenomath*}
\begin{subequations}
\label{S_arr}%
\begin{align}
&  \hat{S}_{\mathrm{gain}}[f^{\prime}]=\sum_{\beta}mm_{\beta}\left(
m+m_{\beta}\right)  ^{2}\int f^{\prime}f_{\beta}^{\prime}\left(
\frac{u^{\prime}}{u}\right)  G\left(  u^{\prime},\Lambda^{\prime}\right)
d^{3}V_{\beta}d^{3}V_{\beta}^{\prime}d^{3}V^{\prime}\nonumber\\
&  \times\delta\left(  m\vec{V}^{\prime}+m_{\beta}\vec{V}_{\beta}^{\prime
}-m\vec{V}-m_{\beta}\vec{V}_{\beta}\right)  \delta\left(  E^{\prime}+E_{\beta
}^{\prime}-E-E_{\beta}-\Delta E_{\mathrm{in}}\right) \label{S_arr(W)}\\
&  =\sum_{\beta}\int f^{\prime}\ \frac{\left(  u^{\prime}\right)  ^{2}}%
{u}\ G\left(  u^{\prime},\Lambda\right)  f_{\beta}^{\prime}d^{3}V_{\beta
}^{\prime}d\Omega_{\mathrm{s}}, \label{S_arr_traditional}%
\end{align}
\end{subequations}
\end{linenomath*}
where $f^{\prime}\equiv f(\vec{V}^{\prime})$, $f_{\beta}^{\prime}\equiv
f_{\beta}(\vec{V}_{\beta}^{\prime})$, and the single integral sign denotes
integrations over all three 3D volumes of $\vec{V}_{\beta}$, $\vec{V}_{\beta
}^{\prime}$, and $\vec{V}^{\prime}$. Equation~(\ref{S_arr}) generalizes the
corresponding part of the Boltzmann collision operator
\citep{Huang:Statistical87,Lifshitz:Physical1981,Shkarofsky:Particle66} to non-elastic collisions. Each of its components
will be defined and explained in the next few pages.

In the general case, the summation over $\beta$ should include all collision
partners described by $f_{\beta}$ or $f_{\beta}^{\prime}$, even the given
particles described by $f$. However, since we neglect collisions between
meteoric particles, the subscript $\beta$ in equations (\ref{D_t(f)}%
)--(\ref{S_arr}) pertains exclusively to the atmospheric particles. This makes
integro-differential equation~(\ref{D_t(f)}) linear with respect to $f$.

The left inequality of equation~(\ref{hierarchy_energies}) allows approximating
atmospheric particles in the meteoroid frame by a monoenergetic beam with the
$\delta$-function velocity distribution,%
\begin{linenomath*}
\begin{equation}
f_{\mathrm{A}}\approx n_{\mathrm{A}}\delta(\vec{V}-\vec{U}).
\label{F_Atm_delta-function}%
\end{equation}
\end{linenomath*}
The last, approximate, equality in equation~(\ref{nu(V)}) uses
equation~(\ref{F_Atm_delta-function}) where $n_{\mathrm{A}}(h(t))$ is the
given local atmospheric density at an altitude $h$ crossed by the meteoroid at
a given moment $t$.

The linear character of kinetic equation~(\ref{D_t(f)}) allows one to treat
each group of colliding particles separately with separate terms like in
equations~(\ref{nu(V)}) and (\ref{S_arr}), each responsible for a given kind
of collision processes. These processes should include all elastic and
inelastic collisions with the atmosphere, e.g., collisional ionization of
meteoric atoms or elastic scattering of the corresponding ions. The ionization
part of the `loss' term for neutrals becomes the corresponding `gain' term for
the newly born ions.

In equation~(\ref{S_arr}), $G\left(  u^{\prime},\Lambda\right)  =d\sigma
/d\Omega_{\mathrm{s}}$ is the differential cross-section taken as a function
of frame-invariant variables, such as $u^{\prime}=|\vec{u}^{\prime}|$ and%
\begin{linenomath*}
\begin{equation}
\Lambda\equiv\cos\Theta_{\mathrm{s}}=\frac{\vec{u}\cdot\vec{u}^{\prime}%
}{uu^{\prime}}. \label{u,Lambda,Phi}%
\end{equation}
\end{linenomath*}
Here $\vec{u}^{\prime}=\vec{V}^{\prime}-\vec{V}_{\beta}^{\prime}$ and $\vec
{u}=\vec{V}-\vec{V}_{\beta}$ are the relative velocities of the two colliding
particles before and after the collision, respectively; in
equation~(\ref{S_arr}) all primes denote particle velocities before the
collision, $\vec{V}^{\prime},\vec{V}_{\beta}^{\prime}\rightarrow\vec{V}%
,\vec{V}_{\beta}$. The scattering angle $\Theta_{\mathrm{s}}$ in
equation~(\ref{u,Lambda,Phi}) is defined as the polar angle of $\vec
{u}^{\prime}$ in the spherical system with the major axis directed along
$\vec{u}$. In the absence of strong external fields, the colliding molecules
are on average unpolarized, hence we assume $G$ to be independent of the
second, i.e., axial, scattering angle, $\Phi_{\mathrm{s}}$. In the general
case of inelastic collisions with the collisional loss of the total kinetic
energy $\Delta E_{\mathrm{in}}$, $u^{\prime}$ and $u$ are related as%
\begin{linenomath*}
\begin{equation}
u^{\prime}=\left(  u^{2}+\frac{2\Delta E_{\mathrm{in}}}{M}\right)  ^{1/2}.
\label{u'^2=u^2-I}%
\end{equation}
\end{linenomath*}
where $M=mm_{\beta}/(m+m_{\beta})$ is the reduced mass of the two colliding particles.

Equation~(\ref{S_arr}) shows two equivalent forms of $\hat{S}_{\mathrm{gain}%
}[f^{\prime}]$. Equation~(\ref{S_arr(W)}) gives a probabilistic form that
precedes the conventional Boltzmann form \citep{Lifshitz:Physical1981} (in the
strict sense, the latter is only applicable to elastic collisions). We will
employ the form (\ref{S_arr(W)}) because it allows for a more accurate account
of inelastic collisions and, more importantly, this form will enable us to
obtain an explicit analytic solution for the velocity distribution of the
secondary particles (see
section~\ref{Velocity distribution of secondary particles}).
Equation~(\ref{S_arr_traditional}) allows us to verify conservation of the
total number of colliding particles,%
\begin{linenomath*}
\begin{equation}
\int\hat{S}_{\mathrm{gain}}[f^{\prime}]d^{3}V^{\prime}=\int\nu(\vec{V}%
)fd^{3}V. \label{local_conservations}%
\end{equation}
\end{linenomath*}
According to equation~(\ref{D_t(f)}), this leads to the stationary continuity
equation:%
\begin{linenomath*}
\begin{equation}
\nabla\cdot\vec{J}=0. \label{continuity}%
\end{equation}
\end{linenomath*}
Equations (\ref{local_conservations}) and (\ref{continuity}) directly apply to
particle scattering without ionization. If inelastic collisions involve
ionization, i.e., a partial conversion of one group (neutral particles) to
another group (ions) then equations~(\ref{local_conservations}) and
(\ref{continuity}) can be easily extended.

In equation~(\ref{S_arr(W)}), the $\delta$-functions in the integrand express
the conservation of the total particle energy (there $E\equiv mV^{2}/2$ and
$E_{\beta}\equiv m_{\beta}V_{\beta}^{2}/2$) and momentum of all colliding
particles during one collision act. In the center-of-mass frame, $m\vec
{V}+m_{\beta}\vec{V}_{\beta}=m\vec{V}^{\prime}+m_{\beta}\vec{V}_{\beta
}^{\prime}=0$, we easily express the individual particle velocities in terms
of $\vec{u}$, $\vec{u}^{\prime}$ as
\begin{linenomath*}
\begin{align}
\vec{V}  &  =\frac{m_{\beta}\vec{u}}{m+m_{\beta}},\qquad\vec{V}_{\beta
}=-\ \frac{m\vec{u}}{m+m_{\beta}},\nonumber\\
\vec{V}^{\prime}  &  =\frac{m_{\beta}\vec{u}^{\prime}}{m+m_{\beta}},\qquad
\vec{V}_{\beta}^{\prime}=-\ \frac{m\vec{u}^{\prime}}{m+m_{\beta}},
\label{VV_COM}%
\end{align}
\end{linenomath*}
so that%
\begin{linenomath*}
\begin{equation}
E+E_{\beta}=\frac{Mu^{2}}{2},\qquad E^{\prime}+E_{\beta}^{\prime}%
=\frac{M(u^{\prime})^{2}}{2}. \label{E+E}%
\end{equation}
\end{linenomath*}
Equation~(\ref{VV_COM}) yields the relation $d^{3}V^{\prime}=[m/(m+m_{\beta
})]^{3}\left(  u^{\prime}\right)  ^{2}du^{\prime}d\Omega_{\mathrm{s}}$, which
helps check the equivalence of the two forms of $\hat{S}_{\mathrm{gain}%
}[f^{\prime}]$. Equation~(\ref{E+E}), along with the energy conservation,
yields equation~(\ref{u'^2=u^2-I}).

Strictly speaking, the integrals in (\ref{D_t(f)}) should also include
summations over discrete energies of the internal degrees of freedom (for
inelastic scattering) and continuous integrations over the kinetic energies of
the released free electron (for ionizing collisions). For simplicity, however,
we consider only one average discrete energy loss, $\Delta E_{\mathrm{in}}$.
For example, for ionizing collisions, we set $\Delta E_{\mathrm{in}%
}=I+\left\langle E_{e}\right\rangle $, where $I$ is the ionization potential
potential and $\left\langle E_{e}\right\rangle $ is the average kinetic energy
of the released electrons. For the present treatment, the imposed inaccuracy
is not important because the corresponding energy losses (usually $<10~$eV)
are small compared to the typical kinetic energies of the colliding heavy
particles, as discussed in
section~\ref{Physical conditions and principal theoretical approximations}. When
obtaining analytic solutions of the kinetic equation, we will neglect $\Delta
E_{\mathrm{in}}$, making inelastic processes essentially equivalent to the
elastic ones. This is even more so for the total momentum changes associated
with the release of a free electron after an ionizing collision. We have not
even included such momentum changes, $P_{\mathrm{in}}$, in the argument of the
corresponding $\delta$-function because the relative momentum changes are much
smaller than the corresponding relative energy changes, $P_{\mathrm{in}%
}/(mV)\lesssim\left(  m_{e}/m\right)  \left[  \Delta E_{\mathrm{in}}%
/(mV^{2}/2)\right]  \ll\Delta E_{\mathrm{in}}/(mV^{2}/2)$.

Collisions of meteor particles with atmospheric molecules have a twofold
effect: they scatter and ionize the meteoroid particles, and at the same time
they scatter the atmospheric molecules and can even ionize the latter, albeit
with a smaller efficiency. On a par with the meteoric particles, the scattered
atmospheric molecules and the corresponding molecular ions could also be
included as separate kinetic groups with the corresponding velocity
distributions. However, the relative number of scattered atmospheric molecules
is so low compared to the total amount of undisturbed atmospheric particles
that they will be of no interest to us. By contrast, the collisionally born
molecular ions of atmospheric origin could in principle make a noticeable
contribution to the near-meteoroid plasma.

Before proceeding with the solution of equation~(\ref{D_t(f)}), we note the
following. If calculated classically, $\nu$ defined by equation~(\ref{nu(V)})
diverges, unless the interaction force between the two colliding particles
becomes exactly zero when the interparticle distance $R$ exceeds a certain
finite value (as in hard-sphere collisions). The `gain' term $\hat
{S}_{\mathrm{gain}}[f^{\prime}]$ has the same problem. The diverging
long-distance part of $G(u,\Lambda)$ corresponds to scattering through small
angles, $\Theta_{s}\rightarrow0$. In terms of equation~(\ref{nu(V)}) and
(\ref{S_arr}), this means a non-integrable singularity of $G(u,\Lambda)$ when
approaching the upper integration limit of $\Lambda=1$. In Coulomb collisions
with the interaction potential $V_{\mathrm{int}}\propto1/R$, the long-distance
scattering through small angles dominates. However, if at least one of the
interacting particles is neutral then the asymptotic long-distance interaction
becomes $V_{\mathrm{int}}\propto1/R^{n}$ with $n\geq4$
\citep{McDaniel:Atomic:Electron89,McDaniel:Atomic:Heavy93,Kaganovich:Scaling06}.
The corresponding scattering through small angles
is no longer dominant and plays insignificant role in collisional changes of
the particle momenta. Nonetheless, even exponentially decreasing interactions
would formally lead to the divergent total cross-section. More accurate
quantum-mechanical calculations yield convergent $\nu$, but they can lead to
operating with physically meaningless long distances.

We will avoid the integral divergence by noticing the following. The formally
dominant contribution\ of small-angle scattering means that $\hat
{S}_{\mathrm{gain}}[f^{\prime}]$ and $\nu(\vec{V})f$ in equation~(\ref{D_t(f)}%
) become indefinitely large, but they must almost perfectly balance each other
to provide a finite difference. In collisions that involve at least one
neutral particle, small-angle scattering creates so minuscule momentum changes
that they result in no perceptible phase-space redistributions of particles.
Then by slightly regrouping the particles we can renormalize $\hat
{S}_{\mathrm{gain}}[f^{\prime}]$ and $\nu(\vec{V})f$ in such a way that would
eliminate the closely balancing large parts. In order to avoid a tedious
renormalization procedure, we will merely assume the given cross-section
$G(u,\Lambda)$ to be a regular function of $\Lambda$. This assumption is
equivalent to cutting off the non-essential asymptotic parts of long-distance
interactions, so that ion-neutral and neutral-neutral collisions become in a
sense similar to hard-sphere collisions.

\section{Outline of the general kinetic
solution\label{Schematics of the general solution}}

To solve equation~(\ref{D_t(f)}), we will use the kinetic approach developed
previously for Compton scattering of $\gamma$-quanta propagating through dense
air \citep{DimantNusinovich:Propagation2012}. We will represent the total
velocity distribution, $f$, as a series of partial distributions,
\begin{linenomath*}
\begin{equation}
f=f^{\left(  1\right)  }+f^{\left(  2\right)  }+f^{(3)}+\ldots,
\label{f_my_approach}%
\end{equation}
\end{linenomath*}
where the superscript denotes the total number of collisions the corresponding
particles experienced since their original ablation plus one. For example,
$f^{\left(  1\right)  }$ describes the sub-group of primary particles ablated
from the meteoroid and freely propagating before they encounter their first
collisions. The function $f^{\left(  2\right)  }$ describes a sub-group of
secondary particles that experienced exactly one collision since their
original ablation. This sub-group includes neutral particles scattered once
and newly-born ions originated during an ionizing collision of a primary
particle with the atmosphere. The function $f^{(3)}$ describes a sub-group of
tertiary particles that experienced exactly two collisions since the original
ablation, and so forth.

The linear character of kinetic equation~(\ref{D_t(f)}) allows one to obtain a
closed differential equation~for the partial function $f^{(j)}$ in terms of
$f^{(j-1)}$,%
\begin{linenomath*}
\begin{equation}
(\vec{V}\cdot\nabla)f^{(j)}+\nu^{(j)}(u)f^{(j)}=\hat{S}_{\mathrm{gain}%
}[f^{(j-1)}], \label{df(J)/dt+nuf}%
\end{equation}
\end{linenomath*}
where the superscripts describe the corresponding particle sub-groups and
$\nu^{(j)}(u)$ is given by equation~(\ref{nu(V)}) for the corresponding
sub-group. Particle conservation requires that the collision terms satisfy the
relationship%
\begin{linenomath*}
\begin{equation}
\int\hat{S}_{\mathrm{gain}}[f^{(j-1)}]d^{3}V=\int\nu^{(j)}(u)f^{(j)}d^{3}V,
\label{Int_S_arr=Int_S_dep}%
\end{equation}
\end{linenomath*}
similar to (\ref{local_conservations}). Equation~(\ref{Int_S_arr=Int_S_dep})
is useful for checking the solutions.

\begin{figure}[h]
\centering
\includegraphics[width=30pc]
{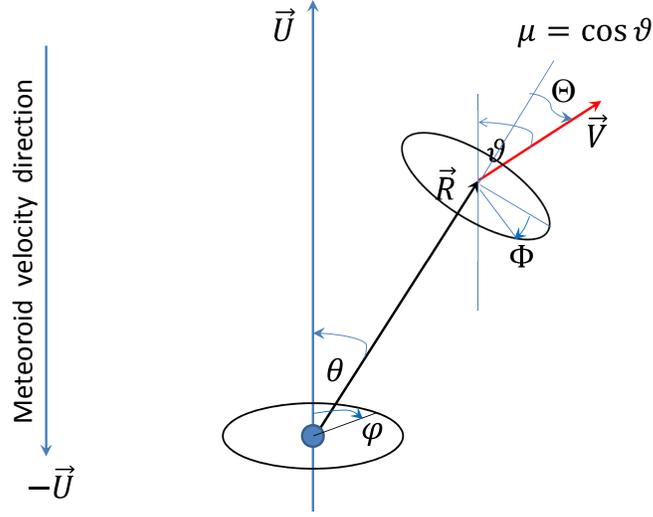}
\caption{Nomenclature of spatial coordinates and velocity variables. The
spatial variables $R=|\vec{R}|$, $\theta$, $\varphi$ denote the radius and two
angles of the spherical coordinate system with the origin at the meteoroid
center and the major axis anti-parallel to the meteoroid velocity shown on the
left. All other variables pertain to the particle velocity space: $V=|\vec
{V}|$ is the particle speed, $\vartheta$ is the polar angle of $\vec{V}$ with
respect to the local axis parallel to $\vec{U}$, $\Phi$ is the axial angle
measured from the common $\vec{U}$-$\vec{R}$ plane; $\Theta$ is the polar
angle of $\vec{V}$ with respect to the local radial distance $\vec{R}$.}%
\label{Fig:Cartoon}
\end{figure}

The only preferred direction in the meteoroid frame is the velocity of the
impinging atmospheric beam, $\vec{U}$. We will represent $\vec{U}$
vertically directed, as in Figure~\ref{Fig:Cartoon}.
In real space, we will use the spherical coordinate system characterized by
the radial distance $R$ measured from the meteoroid center, the polar angle
$\theta$ with respect to the major axis aligned with $\vec{U}$, and the
corresponding axial angle $\varphi$, as shown in Figure~\ref{Fig:Cartoon}.

\begin{figure}[h]
\centering
\includegraphics[width=30pc]
{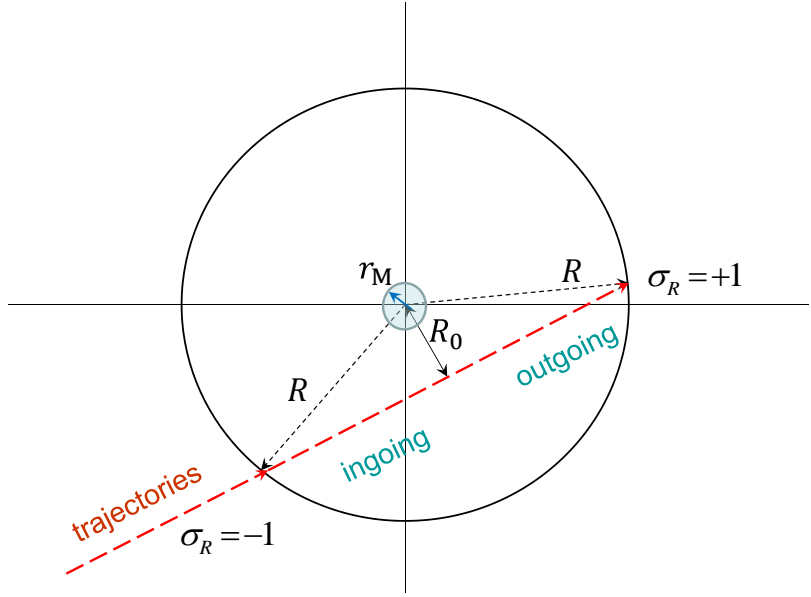}
\caption{A particle trajectory crossing a sphere of a given radial distance
$R$ from the meteoroid center. Here $\varepsilon$ is the meteoroid radius and
$R_{0}$ is the minimum distance from the meteoroid center to the trajectory.
When an incoming particle ($dr/dt<0$, $\sigma_{R}=-1$) crosses $R_{0}$ it
becomes outgoing ($dr/dt>0$, $\sigma_{R}=+1$).}%
\label{Fig:Scheme}
\end{figure}

Equation~(\ref{df(J)/dt+nuf}) with the given RHS can be solved by
characteristics. However, in the velocity space we should use variables that
remain invariant along the ballistic particle trajectories and at the same
time employ the assumed symmetry. We will use variables $V$, $R_{0}$ and
$\Phi$, where $V$ is the particle speed, $R_{0}=R\sin\Theta$, while $\Theta$
and $\Phi$ are respectively the polar and axial angles corresponding to the
local spherical system in the velocity space with the major axis aligned with
$\vec{R}$. Here $\vec{R}$ is the spatial radius-vector measured from the
meteoroid center. The variable $R_{0}$ is the minimum distance between the
straight-line trajectory of a particle and the meteoroid center, as shown in
Figure~\ref{Fig:Scheme}. This variable can also be interpreted as a renormalized
absolute value of the particle angular momentum.

For the primary particles, $j=1$, the RHS of equation~(\ref{df(J)/dt+nuf}) is
zero, so that the solution for $f^{\left(  1\right)  }$ is determined only by
the boundary conditions on the meteoroid surface and the collisional losses
described by $\nu^{(1)}(u)$. For arbitrary $j>1$, the solutions for
$f^{\left(  j\right)  }$ can be obtained recursively, starting from
$f^{\left(  1\right)  }$. The problem is, however, that $\hat{S}%
_{\mathrm{gain}}[f^{(j-1)}]$ defined by equation~(\ref{S_arr(W)}) involves
multiple integrations, so that the mathematical complexity of finding
$f^{\left(  j\right)  }$ increases dramatically with each following $j$.

Fortunately, the physics of plasma formed around a fast moving meteoroid
allows cutting the infinite chain of equation~(\ref{df(J)/dt+nuf}) at a
specific sub-group with $j=k\geq2$. The rest of the particles can be included
in $f^{\left(  k\right)  }$ by dropping the second (loss) term in the LHS of
equation~(\ref{df(J)/dt+nuf}), $\nu^{(j)}f^{(j)}.$ This will automatically
provide the particle conservation. Dropping $\nu^{(j)}f^{(j)}$ will also
simplify the solution. The downside of including $f^{\left(  j\right)  }$ with
$j>k$ to $f^{\left(  k\right)  }$ is that the scattered ions with $j>k$ will
appear closer to the meteor plasma forefront than they actually are. However,
this error will become noticeable only at large distances from the meteoroid,
$R\gtrsim\lambda^{\left(  k\right)  }\gg\lambda^{\left(  1\right)  }$, where
this plasma is mostly inconsequential for the radar head echo.

\begin{figure}[h]
\centering
\includegraphics[width=30pc]
{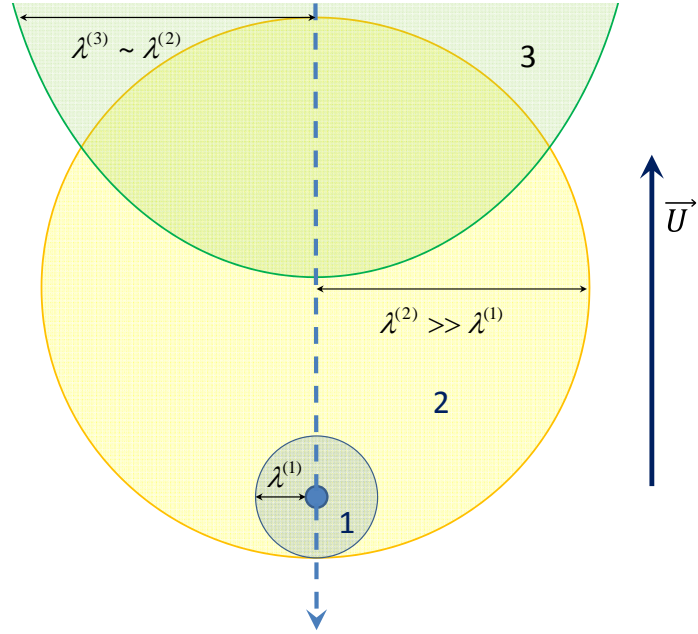}
\caption{A cartoon of core spatial distributions of the primary (1), secondary
(2), and tertiary (3) particles in the meteoroid frame of reference. The core
distributions of the higher-order particle sub-groups are shifted farther away
from the meteoroid in the direction of $\vec{U}$ (i.e., opposite to the
meteoroid velocity). The spherically symmetric core distribution of the
primary particles (with the effective radius $\sim\lambda^{(1)}$) occupies a
much smaller volume than the core distributions of all other sub-groups.
Primary particles are almost entirely neutral, while all other groups include
scattered neutrals and ions.}
\label{Fig:SubGroups}
\end{figure}

Indeed, in the meteoroid frame, particles forming $f^{\left(  1\right)  }$
collide with atmospheric particles that have a huge velocity, $\vec{V}%
\approx\vec{U}$. As a result, the secondary particles described by $f^{(2)}$
have velocities oriented such that their projection onto the $\vec{U}$-line is
almost exclusively in the $\vec{U}$-direction. As $j$ increases further, the
core of the corresponding velocity distribution in the meteoroid frame shifts
closer to $\vec{U}$. In the atmospheric frame, this means that multiply
scattered particles will gradually slow and cool down with the core of the
corresponding particle population in the real space lagging behind the
fast-descending meteoroid and forming the extended trail. Thus the
near-meteoroid sheath is mostly formed by particles with the lowest $j=1,2$,
although a small fraction of each population remains in every location.
Figure~\ref{Fig:SubGroups} shows schematically such spatial distribution. The
distributions $f^{(1,2)}$ should dominate at the forefront of the
near-meteoroid sheath, but behind the meteoroid there is no clear interface
between the near-meteoroid sheath and extended trail. For our purposes,
however, this is unimportant because head-echo radar scattering is determined
mostly by the frontal and side edges of the near-meteoroid plasma and less so
by the elongated trail behind it. Notice that as described below the
characteristic length scale of particle populations with $n\geq2$ is much
larger than that of the primary particles, $\lambda^{(n)}\gg\lambda^{(1)}$.

In this paper, we restrict our kinetic theory to the two lowest-order
functions $f^{(1,2)}$ that describe the primary and secondary particles that
include all higher-order local sub-groups in $f^{(2)}$ through dropping in
equation~(\ref{df(J)/dt+nuf}) the $\nu^{(2)}f^{(2)}$ term. This approximation
will describe the spatial and velocity distribution of all near-meteoroid ions
with sufficient accuracy.

\section{Velocity distribution of the primary
particles\label{Velocity distribution of primary particles}}

If we neglect direct sputtering of the meteoroid by the impinging atmospheric
flow then the velocity distribution of particles ablating from the meteoroid
is fully determined by its structure, material composition, and temperature.
Since we assume a spherical meteoroid with the radius $r_{\mathrm{M}}$ and
uniform temperature $T_{\mathrm{M}}$, the spatial distribution of the primary
particles will also be spherically symmetric. The ablated particles initially
keep their thermal equilibrium, so that on the meteoroid surface,
$R=r_{\mathrm{M}}$, they obey a Maxwellian distribution with the temperature
$T_{\mathrm{M}}$. Since we can linearly combine many sources of the ablating
particles, we consider one species with the mass $M_{\mathrm{met}}$ and given
density $n_{0}$,
\begin{linenomath*}
\begin{equation}
\left.  f^{\left(  1\right)  }\right\vert _{R=r_{\mathrm{M}}}=F_{0}\left(
V\right)  \mathrm{H}\left(  \cos\Theta\right)  ,\qquad F_{0}\left(  V\right)
=\frac{2n_{0}}{\left(  2\pi\right)  ^{3/2}V_{T}^{3}}\mathrm{\ }\exp\left(
-\ \frac{V^{2}}{2V_{T}^{2}}\right)  . \label{F_0(V,Theta)}%
\end{equation}
\end{linenomath*}
Here $V_{T}=(T_{\mathrm{M}}/M_{\mathrm{met}})^{1/2}$ is the thermal velocity
of the ablating gas, $\Theta$ is the polar angle in the velocity space
measured from the local radius-vector $\vec{R}$ on the meteoroid surface, and
$\mathrm{H}\left(  x\right)  $ is the Heaviside step-function ($\mathrm{H}%
\left(  x\right)  =1$ if $x\geq0$ and $\mathrm{H}\left(  x\right)  =0$
otherwise). This step-function expresses the fact that ablating particles
leaving the meteoroid can move only outward from the its surface. The
corresponding radial flux density near the meteoroid surface is $(J_{R}%
)_{0}=\int\left.  f^{\left(  1\right)  }\right\vert _{R=r_{\mathrm{M}}}\left(
V\cos\Theta\right)  d^{3}V=\sqrt{2/\pi}\ n_{0}V_{T}\approx0.8n_{0}V_{T}$. The
parameters $n_{0}$, or $(J_{R})_{0}$, and $T_{\mathrm{M}}$ can be taken from
various ablation models, such as CAMOD \citep{Vondrak:Chemical2008} and others
\citep{Furman:The_theory_I_1960,Furman:A_theory_II_1961,Lebedinets:1970Meteor,
Lebedinets:Interaction1973,Sorasio:Role2001,Mendis:Micrometeoroid2005,Dyrud:Modeling2005,
Campbell-Brown:Meteoroid2007,Berezhnoy:Formation2010}.

At arbitrary distances from the meteoroid center, $R$, the solution for
$f^{\left(  1\right)  }$ can be obtained separately in two overlapped regions:
the closer distances $r_{\mathrm{M}}\leq R\ll\lambda^{\left(  1\right)  }$,
where the mean free path for the primary particles, $\lambda^{\left(
1\right)  }$, is defined by equation~(\ref{lambda^(1),nu}), and the farther
distances, $R\gg r_{\mathrm{M}}$, that include $R\gtrsim\lambda^{\left(
1\right)  }$. In the former region, the primary particles experience
essentially no collisions so that $f^{\left(  1\right)  }$ is given by
equation~(\ref{F_0(V,Theta)}), where the local variable $\Theta$ should be
expressed in terms of the invariant variable $R_{0}=R\sin\Theta$. At farther
distances, $f^{\left(  1\right)  }$ starts experiencing collisional losses,
but the particles propagate mostly along the local radius-vector $\vec{R}$. As
a result, we have to solve equation~(\ref{df(J)/dt+nuf}) without the RHS,
$(\vec{V}\cdot\nabla)f^{\left(  1\right)  }+\nu^{\left(  1\right)
}(u)f^{\left(  1\right)  }=0$ with $\vec{V}\cdot\nabla\approx V(\partial
/\partial R)$. After finding the corresponding solution, matching the two
expressions in the overlap, $r_{\mathrm{M}}\ll R\ll\lambda^{\left(  1\right)
}$, and expressing the invariant variable $R_{0}$ back in terms of the local
variable $\Theta$ for the given location $\vec{R}$, we obtain%
\begin{linenomath*}
\begin{equation}
f^{\left(  1\right)  }\approx\frac{2n_{0}}{\left(  2\pi\right)  ^{3/2}%
V_{T}^{3}}\ \mathrm{H}\left(  \cos\Theta-\sqrt{1-\frac{r_{\mathrm{M}}^{2}%
}{R^{2}}}\right)  \exp\left(  -\ \frac{V^{2}}{2V_{T}^{2}}-\frac{R}%
{\lambda^{\left(  1\right)  }}\right)  , \label{f^(0)_general}%
\end{equation}
\end{linenomath*}
where the mean free path for the primary particles, $\lambda^{\left(
1\right)  }$, is given by%
\begin{linenomath*}
\begin{equation}
\lambda^{\left(  1\right)  }(V)=\frac{V}{\nu^{\left(  1\right)  }},\qquad
\nu^{\left(  1\right)  }\approx2\pi n_{\mathrm{A}}U\int_{-1}^{1}G\left(
U,\Lambda\right)  d\Lambda. \label{lambda^(1),nu}%
\end{equation}
\end{linenomath*}
Here we have taken into account the fact that $V\sim V_{T}\ll U$, so that
$u\approx U$. Equation (\ref{lambda^(1),nu}) shows that the collision
frequency $\nu^{\left(  1\right)  }$ is approximately constant and
$\lambda^{\left(  1\right)  }\propto V$, i.e., faster primary particles move
farther away from the meteoroid before being collisionally scattered or
ionized. Since $V\ll U$, the mean free path of the primary particles,
$\lambda^{\left(  1\right)  }(V)$, is much shorter than the mean free paths
for all other sub-groups with $V\sim U$, but is still many orders of magnitude
longer than the meteoroid radius, $r_{\mathrm{M}}$.

\section{Velocity distribution of the secondary
particles\label{Velocity distribution of secondary particles}}

The group of secondary particles, $f^{\left(  2\right)  }$, includes most ions
of the near-meteoroid plasma. Calculation of $f^{\left(  2\right)  }$ is more
complicated than $f^{\left(  1\right)  }$ and is the central topic of this paper.

After cutting the infinite chain described by equation~(\ref{df(J)/dt+nuf}) at
$k=2$ and dropping the corresponding loss term, the closed kinetic
equation~for $f^{\left(  2\right)  }$ becomes
\begin{linenomath*}
\begin{equation}
(\vec{V}\cdot\nabla)f^{(2)}=\hat{S}_{\mathrm{gain}}[f^{(1)}]. \label{df^(2)}%
\end{equation}
\end{linenomath*}
In order to obtain the explicit analytical expression for $f^{\left(
2\right)  }$, we have to calculate $\hat{S}_{\mathrm{gain}}[f^{\left(
1\right)  }]$ and then integrate it over the characteristics of
equation~(\ref{df^(2)}). Before proceeding, we will first discuss individual
collisions of primary particles with an almost monoenergetic beam of
atmospheric molecules. This analysis will suggest a relevant approximation of
$\hat{S}_{\mathrm{gain}}[f^{(1)}]$ that will eventually lead us to a proper
analytic solution for $f^{\left(  2\right)  }$.

\subsection{Collisions of individual primary particles with the
atmosphere\label{Collisions of individual primary particles with the atmosphere}%
}

Consider an individual collision of a primary meteoric particle with a cold
atmospheric particle. Before the collision, the meteoric-particle velocity in
the meteoroid frame was $\vec{V}^{\prime}$, while the atmospheric molecule
velocity in the same frame was $\vec{V}_{\beta}^{\prime}=\vec{U}$. Given the
meteoric particle velocity after the collision, $\vec{V}$, and applying the
momentum conservation, we obtain the atmospheric-molecule velocity immediately
after the collision:
\begin{linenomath*}
\begin{equation}
\vec{V}_{\beta}=\frac{m}{m_{\beta}}(\vec{V}^{\prime}-\vec{q}),\qquad\vec
{q}\equiv\vec{V}-\frac{m_{\beta}\vec{U}}{m} \label{V_beta_sec}%
\end{equation}
\end{linenomath*}
(the reason for introducing $\vec{q}$ will become clear in
section~\ref{Calculation of S_arr}). From this point on, we will neglect
inelastic losses of energy, so that $u=u^{\prime}$. Rewriting the latter as
$u^{2}-\left(  u^{\prime}\right)  ^{2}=\left(  \vec{u}+\vec{u}^{\prime
}\right)  \cdot\left(  \vec{u}-\vec{u}^{\prime}\right)  =0$ with $\vec
{u}\equiv\vec{V}-\vec{V}_{\beta}$, $\vec{u}^{\prime}\equiv\vec{V}^{\prime
}-\vec{V}_{\beta}^{\prime}=\vec{V}^{\prime}-\vec{U}$, and using
equation~(\ref{V_beta_sec}) for $\vec{V}_{\beta}$, we obtain%
\begin{linenomath*}
\begin{equation}
\left[  \left(  1+\frac{m}{m_{\beta}}\right)  \vec{V}+\left(  1-\frac
{m}{m_{\beta}}\right)  \vec{V}^{\prime}-2\vec{U}\right]  \cdot(\vec{V}-\vec
{V}^{\prime})=0. \label{exact_via_U_0}%
\end{equation}
\end{linenomath*}
As discussed above, before the collision a typical primary-particle speed,
$V^{\prime}\sim(2T_{\mathsf{M}}/m_{\mathrm{M}})^{1/2}$, was small compared to
$U$. Then to the zeroth-order accuracy corresponding to $\vec{V}^{\prime
}\rightarrow0$ equation~(\ref{exact_via_U_0}) yields
\begin{linenomath*}
\begin{equation}
V\approx V_{Q}(\mu)\equiv\frac{2\mu m_{\beta}U}{m+m_{\beta}}, \label{V_Q}%
\end{equation}
\end{linenomath*}
where $\mu\equiv\cos\vartheta=\vec{V}\cdot\vec{U}/(VU)$ and $\vartheta$ is the
polar angle of the particle velocity with respect to the $\vec{U}$ direction,
as shown in Figure~\ref{Fig:Cartoon}. According to equation~(\ref{V_Q}), the
$\vec{V}^{\prime}\rightarrow0$ approximation allows only positive values of
$\mu$: $0\leq\mu\leq1$ ($\pi/2\geq\vartheta\geq0$). Under this approximation,
the relative velocities of the colliding particles and the cosine of the
corresponding scattering angle, $\Lambda=\cos\Theta_{\mathrm{s}}$, are given
by%
\begin{linenomath*}
\begin{align}
\vec{u}^{\prime}  &  \approx-\vec{U},\qquad\vec{u}\approx\left(  1+\frac
{m}{m_{\beta}}\right)  \vec{V}-\vec{U},\nonumber\\
u^{\prime}  &  =u\approx U,\qquad\Lambda=\frac{\vec{u}^{\prime}\cdot\vec{u}%
}{u^{2}}\approx1-2\mu^{2}. \label{Lambda=1-2mu^2}%
\end{align}
\end{linenomath*}
Unlike $\mu$, the scattering parameter $\Lambda$ spans the entire domain
between $1$ and $-1$ ($0\leq\Theta_{\mathrm{s}}\leq\pi$).

The speed of the secondary particles, $V$, reaches its maximum value,
$V_{\max}\approx2m_{\beta}U/\left(  m+m_{\beta}\right)  $, for $\mu=1$
corresponding to the backward scatter, $\Theta_{\mathrm{s}}=\pi$ ($\Lambda
=-1$). In the opposite limit of small-angle scattering, $\Theta_{\mathrm{s}%
}\rightarrow0$ ($\Lambda\rightarrow1$), equations~(\ref{V_Q}) and
(\ref{Lambda=1-2mu^2}) yield $\mu,V\rightarrow0$. In this limit, however, the
approximation of $\vec{V}^{\prime}\rightarrow0$\ is inaccurate. In reality,
particles scattered through small angles acquire, after a collision, not a
zero but a small speed, $V\sim V^{\prime}\ll U$, with arbitrary $\vartheta$.
Equation~(\ref{exact_via_U_0}) accounts for all cases, but fully neglecting
$\vec{V}^{\prime}$ does not work if $\Theta_{\mathrm{s}}\lesssim
\lbrack(m+m_{\beta})/(2m_{\beta})]V^{\prime}/U$ when $V$ and $V^{\prime}$
become comparable. However, under conditions when slow primary particles
collide with extremely fast-moving atmospheric particles, the small-angle
collisions retaining the slow speed of the primary particles, $V\sim
V^{\prime}\ll U$, are so rare that they make no tangible contribution to the
total velocity distribution of the secondary particles. This allows one to
neglect small-angle scattering and employ equations~(\ref{V_Q}%
)--(\ref{Lambda=1-2mu^2}) everywhere, regardless of the inaccuracy near
$\Lambda=1$ ($\mu=0$).

\begin{figure}[h]%
\centering
\includegraphics[width=30pc]
{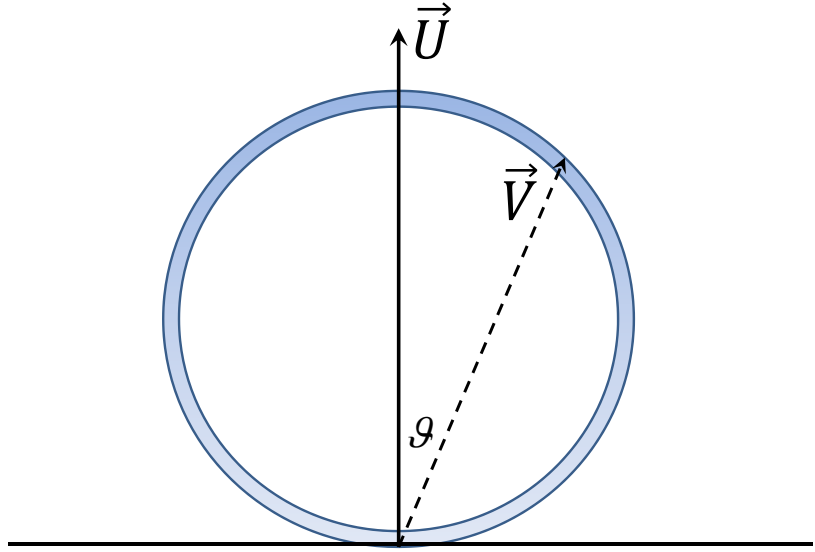}
\caption{Shell-like structure of $f^{(2)}$ (schematic view of a meridional
cross-section of the spherical shell).}
\label{Fig:Shell}
\end{figure}

The fact that the zero-order approximation with respect to $V^{\prime}/U\ll1$
yields the one-to-one correspondence between $V$ and $\mu$ given by
equation~(\ref{V_Q}) leads to the following important consequence. The 3-D
velocity distribution of the secondary particles is concentrated within a thin
spherical shell around $V=V_{Q}(\mu)$, as illustrated schematically by
Figure~\ref{Fig:Shell}. The corresponding sphere is shifted in the $\vec{U}%
$-direction by its radius, $V_{\max}/2=m_{\beta}U/(m+m_{\beta})$. As discussed
above, near the lowest edge where the shell is tangential to the plane
$V_{z}\equiv\mu V=0$, the entire approximation leading to equation~(\ref{V_Q})
is inaccurate. However, the contribution of this edge to the entire velocity
distribution is so small that we will ignore it. The function $f^{(2)}(\vec
{V})$ is non-uniformly distributed over the spherical shell, but for the
moment this is of no importance.

In section~\ref{Calculation of S_arr} below, we will need a better accuracy than
equation~(\ref{V_Q}) provides. Linearizing equation~(\ref{exact_via_U_0}) over
$\vec{V}^{\prime}$, to first-order accuracy with respect to $V^{\prime}/U\ll
1$, we obtain%
\begin{linenomath*}
\[
2\vec{U}\cdot\vec{V}-\left(  1+\frac{m}{m_{\beta}}\right)  V^{2}+2\left(
\frac{m\vec{V}}{m_{\beta}}-\vec{U}\right)  \cdot\vec{V}^{\prime}\approx0.
\]
\end{linenomath*}
This relation allows expressing small deviations of $V$ from $V_{Q}(\mu)$ in
terms of the relatively small velocities $\vec{V}^{\prime}$,%
\begin{linenomath*}
\begin{equation}
V-V_{Q}(\mu)=\frac{m(\vec{q}\cdot\vec{V}^{\prime})}{m_{\beta}\mu U},
\label{q.V'}%
\end{equation}
\end{linenomath*}
where $\vec{q}$ is defined in equation~(\ref{V_beta_sec}).

Using these findings, we will construct an accurate approximation for $\hat
{S}_{\mathrm{gain}}[f^{\left(  1\right)  }]$ given below by
equation~(\ref{S_arr_via_K}) that will dramatically simplify the solution for
$f^{(2)}$.

\subsection{Calculation of $\hat{S}_{\mathrm{gain}}[f^{\left(  1\right)  }%
]$\label{Calculation of S_arr}}

In order to calculate the RHS of equation~(\ref{df^(2)}) in the employed
elastic approximation, $u^{\prime}=u$, we will use for $\hat{S}_{\mathrm{gain}%
}[f^{\left(  1\right)  }]$\ equation~(\ref{S_arr(W)}) with $\Delta
E_{\mathrm{in}}=0$. This unconventional form of the collisional operator will
allow us to avoid complexities associated with expressing the angular
arguments of $f^{\left(  1\right)  }$ in terms of the two scattering angles,
$\Theta_{\mathrm{s}}$ and $\Phi_{\mathrm{s}}$. We will calculate $\hat
{S}_{\mathrm{gain}}[f^{\left(  1\right)  }]$ at a given location,
characterized by the radius-vector $\vec{R}$, using local velocity variables
convenient for the integrations but not necessarily invariants of the
collisionless motion.

After integration over $d^{3}V_{\beta}$ and elimination of the momentum
conservation $\delta$-function, equation~(\ref{S_arr(W)}) reduces to
\begin{linenomath*}
\begin{align*}
&  \hat{S}_{\mathrm{gain}}[f^{\left(  1\right)  }]=n_{\mathrm{A}}m\left(
1+\frac{m}{m_{\beta}}\right)  ^{2}\int f^{\left(  1\right)  }(V^{\prime
},\Omega^{\prime})\\
&  \times G\left(  u,\Lambda\right)  \delta\left(  E^{\prime}+\frac{m_{\beta
}U^{2}}{2}-E-E_{\beta}\right)  \left(  V^{\prime}\right)  ^{2}dV^{\prime
}d\Omega^{\prime},
\end{align*}
\end{linenomath*}
where $\Omega^{\prime}$ is the solid angle in the $\vec{V}^{\prime}$-space.

In this velocity space, we introduce an \emph{ad hoc} spherical system with
the major axis parallel to $\vec{R}$. The corresponding variables
characterizing $\vec{V}^{\prime}$ are $V^{\prime}$, $\Theta^{\prime}$,
$\Phi^{\prime}$, where $\Theta^{\prime}$ and $\Phi^{\prime}$ are the polar and
axial angles respectively, so that $d\Omega^{\prime}=d(\cos\Theta^{\prime
})d\Phi^{\prime}$. To eliminate the remaining $\delta$-function, instead of a
seemingly natural integration over $V^{\prime}$, we will first integrate over
$\Phi^{\prime}$ (the benefit of this integration will become clear later).
This yields%
\begin{linenomath*}
\begin{equation}
\hat{S}_{\mathrm{gain}}[f^{\left(  1\right)  }]=mn_{\mathrm{A}}\left(
1+\frac{m}{m_{\beta}}\right)  ^{2}\sum_{i}\int\frac{f^{\left(  1\right)
}(V^{\prime},\Omega^{\prime})G\left(  u,\Lambda\right)  \left(  V^{\prime
}\right)  ^{2}dV^{\prime}d(\cos\Theta^{\prime})}{\left\vert \partial E_{\beta
}/\partial\Phi^{\prime}\right\vert _{\Phi^{\prime}=\Phi_{i}^{\prime}}},
\label{tushka}%
\end{equation}
\end{linenomath*}
where the $i$-summation is over all roots of the energy-conservation equation
$E^{\prime}+m_{\beta}U^{2}/2-E-E_{\beta}\left(  \Phi^{\prime}\right)  =0$ and
the derivatives $\partial E_{\beta}/\partial\Phi^{\prime}$ should be
calculated before applying this energy conservation. To express $E_{\beta}$ in
terms of $\Phi^{\prime}$, we use equation~(\ref{V_beta_sec}) that has followed
only from the momentum conservation. This yields%
\begin{linenomath*}
\begin{equation}
E_{\beta}=\frac{m\vec{V}_{\beta}^{2}}{2}=\frac{m^{2}}{m_{\beta}}\left[
\frac{(\vec{V}^{\prime})^{2}}{2}-\vec{V}^{\prime}\cdot\vec{q}+\frac{q^{2}}%
{2}\right]  ,\qquad q\equiv\left\vert \vec{q}\right\vert , \label{E_beta}%
\end{equation}
\end{linenomath*}
where the only $\Phi^{\prime}$-dependent term is $-\vec{V}^{\prime}\cdot
\vec{q}$. To find the $\Phi^{\prime}$-dependence, we express $\vec{V}^{\prime
}=(\hat{e}_{R}\cos\Theta^{\prime}+\hat{e}_{\Theta}\sin\Theta^{\prime}\cos
\Phi^{\prime}+\hat{e}_{\Phi}\sin\Theta^{\prime}\sin\Phi^{\prime})V^{\prime}$
and $\vec{q}=(\hat{e}_{R}\cos\Theta_{q}+\hat{e}_{\Theta}\sin\Theta_{q}\cos
\Phi_{q}+\hat{e}_{\Phi}\sin\Theta_{q}\sin\Phi_{q})q$, where $\hat{e}%
_{R,\Theta,\Phi}$ are the mutually orthogonal base vectors (each unit vector
is in the direction of the corresponding coordinate variation). Denoting the
angle between $\vec{V}^{\prime}$ and $\vec{q}$ as $\Psi$, we obtain
\begin{linenomath*}
\begin{equation}
\cos\Psi=\frac{\vec{V}^{\prime}\cdot\vec{q}}{V^{\prime}q}=\cos\Theta^{\prime
}\cos\Theta_{q}+\sin\Theta^{\prime}\sin\Theta_{q}\cos(\Phi^{\prime}-\Phi_{q}).
\label{cos_Psi_opiat}%
\end{equation}
\end{linenomath*}
Equations~(\ref{E_beta}) and (\ref{cos_Psi_opiat}) yield $\left\vert \partial
E_{\beta}/\partial\Phi^{\prime}\right\vert =\left(  m^{2}/m_{\beta}\right)
|\partial(\vec{V}^{\prime}\cdot\vec{q})/\partial\Phi^{\prime}|=\left(
m^{2}V^{\prime}q/m_{\beta}\right)  \left\vert \sin\Theta^{\prime}\sin
\Theta_{q}\sin(\Phi^{\prime}-\Phi_{q})\right\vert $. The energy-conservation
equation, $E^{\prime}+m_{\beta}U^{2}/2-E-E_{\beta}(\Phi_{i}^{\prime})=0$, has
exactly two roots $\Phi_{i}^{\prime}$, $i=1,2$, whose specific values are
inconsequential for determining $\left\vert \partial E_{\beta}/\partial
\Phi^{\prime}\right\vert _{\Phi^{\prime}=\Phi_{i}^{\prime}}$. Then using
equation~(\ref{cos_Psi_opiat}), we obtain%
\begin{linenomath*}
\begin{equation}
\sum_{i=1}^{2}\frac{\left(  \cdots\right)  }{\left\vert \partial E_{\beta
}/\partial\Phi^{\prime}\right\vert _{\Phi^{\prime}=\Phi_{i}^{\prime}}}%
=\frac{2m_{\beta}\left(  \cdots\right)  }{m^{2}V^{\prime}q\sqrt{S}},
\label{sumka}%
\end{equation}
\end{linenomath*}
where the factor `$2$' has originated from the summation over the two roots
$\Phi_{i}^{\prime}$ and the function $S$ is given by
\begin{linenomath*}
\begin{align}
&  S\equiv\left\vert \sin\Theta^{\prime}\sin\Theta_{q}\sin(\Phi^{\prime}%
-\Phi_{q})\right\vert ^{2}\nonumber\\
&  =\left[  \cos(\Theta_{q}-\Theta^{\prime})-\cos\Psi\right]  \times\left[
\cos\Psi-\cos(\Theta_{q}+\Theta^{\prime})\right]  . \label{S_2}%
\end{align}
\end{linenomath*}
Combining equations~(\ref{tushka}) and (\ref{sumka}), we obtain%
\begin{linenomath*}
\begin{equation}
\hat{S}_{\mathrm{gain}}[f^{\left(  1\right)  }]=\frac{2\left(  m+m_{\beta
}\right)  ^{2}n_{\mathrm{A}}}{mm_{\beta}q}\iint\frac{f^{\left(  1\right)
}(V^{\prime},\Omega^{\prime})G(u,\Lambda)}{\sqrt{S}}\ V^{\prime}dV^{\prime
}d(\cos\Theta^{\prime}). \label{prome_S_arr}%
\end{equation}
\end{linenomath*}

\begin{figure}[h]
\centering
\includegraphics[width=30pc]
{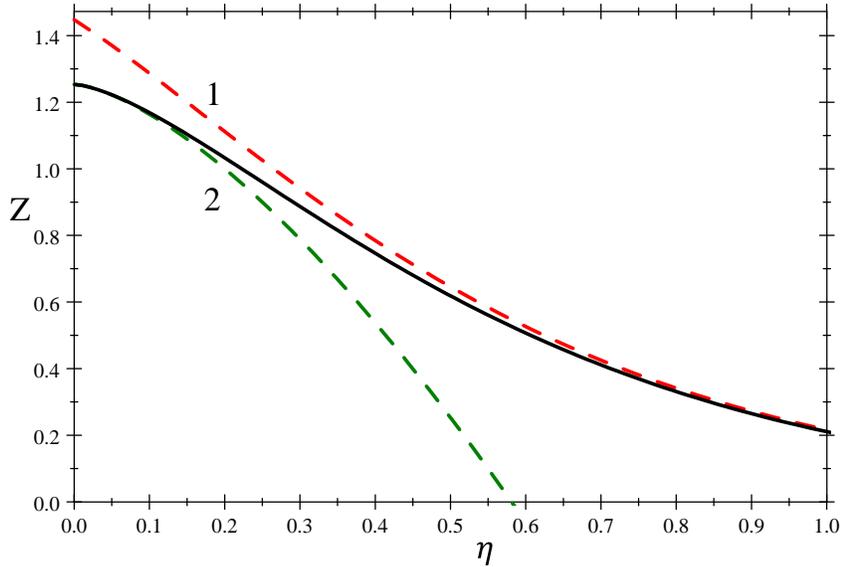}
\caption{Function $Z(\eta)$ defined by equation~(\ref{Z(eta)}). The solid
curve shows the result of exact computer integration, while the dashed curves
show the corresponding large-$\eta$ and small-$\eta$ asymptotics: (1) $\left.
Z\right\vert _{\eta\gg1}$ given by equation~(\ref{Z(V,R)}) and (2) $\left.
Z\right\vert _{\eta\ll1}\approx\sqrt{\pi/2}-2\sqrt{2}\ \eta^{\frac{3}{2}}$.}%
\label{Fig:Z_eta}
\end{figure}

Until this point, all expressions related to $\hat{S}_{\mathrm{gain}%
}[f^{\left(  1\right)  }]$ were exact and have not used the smallness of
$V^{\prime}/U$ discussed in
section~\ref{Collisions of individual primary particles with the atmosphere},
but now we will use it. To the zero-order accuracy with respect to $V^{\prime
}/U$, according to equation~(\ref{Lambda=1-2mu^2}), we have $u\approx U$ and
$\Lambda=\cos\Theta_{\mathrm{sc}}\approx1-2\mu^{2}$. To the same zeroth order,
we have the approximate one-to-one correspondence between $V$ and $\mu$
described by equation~(\ref{V_Q}). This correspondence means that the
collisional source of secondary particles, $\hat{S}_{\mathrm{gain}}[f^{\left(
1\right)  }]$, has an approximate $\delta$-function dependence,%
\begin{linenomath*}
\begin{equation}
\hat{S}_{\mathrm{gain}}[f^{\left(  1\right)  }]\approx K\delta\left(
V-V_{Q}(\mu)\right)  ,\qquad K=\int_{0}^{\infty}\hat{S}_{\mathrm{gain}%
}[f^{\left(  1\right)  }]dV, \label{S_arr_via_K}%
\end{equation}
\end{linenomath*}
that corresponds to an infinitely thin shell distribution discussed in
section~\ref{Collisions of individual primary particles with the atmosphere}%
\ and illustrated in Figure~\ref{Fig:Shell}. However, calculating the factor
$K$ requires better accuracy. According to equations~(\ref{prome_S_arr}) and
(\ref{S_arr_via_K}), we have%
\begin{linenomath*}
\begin{align}
&  K\approx\frac{2\left(  m+m_{\beta}\right)  ^{2}n_{\mathrm{A}}G\left(
U,1-2\mu^{2}\right)  }{mm_{\beta}q}\int_{0}^{\infty}dV\iint\frac{f^{\left(
1\right)  }}{\sqrt{S}}\ V^{\prime}dV^{\prime}d\left(  \cos\Theta^{\prime
}\right) \nonumber\\
&  =\frac{4\left(  m+m_{\beta}\right)  ^{2}n_{\mathrm{A}}n_{0}G\left(
U,1-2\mu^{2}\right)  }{\left(  2\pi\right)  ^{3/2}mm_{\beta}qV_{T}^{3}%
}\label{K}\\
&  \!\times\!\!\int_{0}^{\infty}\!\!dV\!\!\iint\mathrm{H}\!\left(  \cos
\Theta^{\prime}-\sqrt{1-\frac{r_{\mathrm{M}}^{2}}{R^{2}}}\right)
\exp\!\left[  -\ \frac{\left(  V^{\prime}\right)  ^{2}}{2V_{T}^{2}}-\frac
{\nu^{\left(  1\right)  }R}{V^{\prime}}\right]  \!\frac{V^{\prime}dV^{\prime
}d\left(  \cos\Theta^{\prime}\right)  }{\sqrt{S}}.\nonumber
\end{align}
\end{linenomath*}
To proceed with the triple integration, we have to first relate $V$ and
$\cos\Psi$ defined by equation~(\ref{cos_Psi_opiat}) and contained in the
expression for $S$, as defined by equation~(\ref{S_2}). Only this step
requires the first-order expansion with respect to small $V^{\prime}/U$; for
anything else it suffices to use the zero-order relation of $V=V_{Q}(\mu)$.
Given $\vec{q}$, equation~(\ref{q.V'}) relates the speed difference
$(V-V_{Q})$ to the primary-particle velocities $\vec{V}^{\prime}$:%
\begin{linenomath*}
\begin{equation}
\cos\Psi=\frac{\vec{q}\cdot\vec{V}^{\prime}}{qV^{\prime}}\approx\frac
{m_{\beta}\mu\left(  V-V_{Q}\right)  U}{mqV^{\prime}}, \label{cos_Psi}%
\end{equation}
\end{linenomath*}
so that $dV=[mqV^{\prime}/(m_{\beta}\mu U)]d\left(  \cos\Psi\right)  $. This
yields an interim expression for $K$:%
\begin{linenomath*}
\begin{align*}
&  K\approx\frac{4n_{\mathrm{A}}n_{0}G\left(  U,1-2\mu^{2}\right)  }{\left(
2\pi\right)  ^{3/2}\mu V_{T}^{3}U}\left(  1+\frac{m}{m_{\beta}}\right)  ^{2}\\
&  \times\int_{0}^{\infty}\exp\left(  -\ \frac{\left(  V^{\prime}\right)
^{2}}{2V_{T}^{2}}-\frac{\nu^{\left(  1\right)  }R}{V^{\prime}}\right)  \left(
V^{\prime}\right)  ^{2}dV^{\prime}\\
&  \times\iint\mathrm{H}\left(  \cos\Theta^{\prime}-\sqrt{1-\frac
{r_{\mathrm{M}}^{2}}{R^{2}}}\right)  \frac{d(\cos\Theta^{\prime})d(\cos\Psi
)}{\sqrt{S}},
\end{align*}
\end{linenomath*}
where the double integration should be performed over the entire 2-D area of
positive $S$. First, we integrate over $\cos\Psi$. Using equation~(\ref{S_2}),
we obtain $\left\vert \int_{\cos\left(  \Theta_{q}+\Theta^{\prime}\right)
}^{\cos\left(  \Theta_{q}-\Theta^{\prime}\right)  }d\left(  \cos\Psi\right)
/\sqrt{S}\right\vert =\pi$. This exact relation holds even if the two
integration limits are infinitesimally close to each other ($\left\vert
\Theta^{\prime}\right\vert \rightarrow0$). Then, integrating over $\cos
\Theta^{\prime}$, we obtain%
\begin{linenomath*}
\begin{align}
K(\mu,R)  &  \approx\sqrt{\frac{2}{\pi}}\ \frac{n_{\mathrm{A}}n_{0}G\left(
U,1-2\mu^{2}\right)  }{\mu U}\left(  1+\frac{m}{m_{\beta}}\right)  ^{2}\left(
1-\sqrt{1-\frac{r_{\mathrm{M}}^{2}}{R^{2}}}\right)  Z(\eta)\nonumber\\
&  \approx\frac{2n_{\mathrm{A}}n_{0}G\left(  U,1-2\mu^{2}\right)  }{\sqrt
{3}\ \mu U}\left(  1+\frac{m}{m_{\beta}}\right)  ^{2}\left[  1+\left(
\frac{\nu^{\left(  1\right)  }R}{V_{T}}\right)  ^{2/3}\right] \nonumber\\
&  \times\left(  1-\sqrt{1-\frac{r_{\mathrm{M}}^{2}}{R^{2}}}\right)
\exp\left[  -\ \frac{3}{2}\left(  \frac{\nu^{\left(  1\right)  }R}{V_{T}%
}\right)  ^{2/3}\right]  , \label{K(mu,R)}%
\end{align}
\end{linenomath*}
where in the latter, approximate, equality the dimensionless integral,%
\begin{linenomath*}
\begin{align}
&  Z(\eta)\equiv\frac{1}{V_{T}^{3}}\int_{0}^{\infty}\exp\left(  -\ \frac
{V^{2}}{2V_{T}^{2}}-\frac{\nu^{\left(  1\right)  }R}{V}\right)  V^{2}%
dV\label{Z(eta)}\\
&  =\left(  2\eta\right)  ^{\frac{3}{2}}\int_{0}^{\infty}\exp\left[
-\eta\left(  x^{2}+\frac{2}{x}\right)  \right]  x^{2}dx,\qquad\eta=\frac{1}%
{2}\left(  \frac{\nu^{\left(  1\right)  }R}{V_{T}}\right)  ^{\frac{2}{3}%
},\nonumber
\end{align}
\end{linenomath*}
was approximated by its $R\gg\lambda^{(1)}$ ($\eta\gg1$) asymptotic
expression,%
\begin{linenomath*}
\begin{equation}
\left.  Z\right\vert _{\eta\gg1}\approx\sqrt{\frac{2\pi}{3}}\left(
1+2\eta\right)  e^{-3\eta}. \label{Z(V,R)}%
\end{equation}
\end{linenomath*}
Approximate equation~(\ref{Z(V,R)}) works reasonably well for all distances
from the meteoroid center, even in the worst case of $R\ll\lambda_{T}^{(1)}$
when approximate $Z(\eta)$ exceeds the exact expression by a factor of
$2/\sqrt{3}\approx1.155$, as shown in Figure~\ref{Fig:Z_eta}.

Thus the `gain' term for the secondary particles, $\hat{S}_{\mathrm{gain}%
}[f^{\left(  1\right)  }]$, is given by equation~(\ref{S_arr_via_K}), where
the factor $K(\mu,R)$ is given by equation~(\ref{K(mu,R)}). One can verify
that $\hat{S}_{\mathrm{gain}}[f^{\left(  1\right)  }]$ satisfies
equation~(\ref{Int_S_arr=Int_S_dep}). Since $V$ and $\mu$ are invariants of
the particle collisionless motion, the approximate $\delta$-function speed
dependence of equation~(\ref{S_arr_via_K}) translates to a similar $V,\mu
$-dependence for $f^{\left(  2\right)  }$, as described in
equation~(\ref{f^(2)_approximate}) below.

\subsection{Integration over particle
trajectories\label{Integration over particle trajectories}}

Having calculated the RHS of equation (\ref{df^(2)}), we are ready to solve
it~by characteristics. Because the LHS of equation~(\ref{df^(2)}) contains
spatial derivatives we should use the coordinate system appropriate for the
assumed axial symmetry around the direction of $\vec{U}$ and employ invariant
variables in the velocity space. The RHS of (\ref{df^(2)}) depends only on one
spatial coordinate, $R$, so it is convenient to use for the real space the
spherical coordinate system: $R$, $\theta$, and $\phi$, where $\theta$ is the
polar angle with respect to the $\vec{U}$-direction and $\phi$ is the
corresponding axial angle (due to the axial symmetry, there will be no $\phi
$-dependence). Positive values of $\cos\theta=(\vec{R}\cdot\vec{U})/(RU)$
denote locations behind the descending meteoroid, while negative $\cos\theta$
denote locations in front of it. Similar to
section~\ref{Velocity distribution of primary particles}, we will use the
minimum distance of a given straight-line particle trajectory to the meteoroid
center, $R_{0}=R\sin\Theta$, as one of the velocity-space variables that
remain invariant during the particle free motion, see Figure~\ref{Fig:Scheme}.
The entire set of these invariant variables includes $V$, $R_{0}$, and $\Phi$,
where the axial angle $\Phi$ is the axial angle of $\vec{V}$ measured around
the direction of $\vec{R}$ from the common $\vec{R}$-$\vec{U}$ plane. We
should bear in mind that the polar angle in the velocity space, $\Theta$,
measured from the local direction of $\vec{R}$ is not an invariant of the free
particle motion.

In these variables, using the approximation given by
equation~(\ref{S_arr_via_K}), we reduce equation~(\ref{df^(2)}) without the
$\nu^{(2)}f^{\left(  2\right)  }$ term to
\begin{linenomath*}
\begin{equation}
V_{R}(R,V)\ \frac{df^{\left(  2\right)  }}{dR}=K\left(  \mu,R\right)
\delta\left(  V-V_{Q}\left(  \mu\right)  \right)  , \label{start}%
\end{equation}
\end{linenomath*}
where $V_{Q}\left(  \mu\right)  $ is defined by equation~(\ref{V_Q}) and
$K\left(  \mu,R\right)  $ by equation~(\ref{K(mu,R)}). In the LHS of
equation~(\ref{start}), $d/dR$ denotes the full derivative along a given
particle trajectory characterized by $V$, $R_{0}$, $\Phi$, with the
$R$-dependent local radial component of the particle velocity, $V_{R}$, given
by
\begin{linenomath*}
\begin{equation}
V_{R}\equiv V\cos\Theta=\sigma_{R}\sqrt{1-\frac{R_{0}^{2}}{R^{2}}}%
\ V,\qquad\sigma_{R}=\pm1. \label{V_r_snova}%
\end{equation}
\end{linenomath*}
Here $\sigma_{R}$ is an additional discrete parameter which identifies either
`outgoing' ($dR/dt>0$, $\sigma_{R}=+1$) or `incoming' ($dR/dt<0$, $\sigma
_{R}=-1$) particles, as depicted in Figure~\ref{Fig:Scheme}. The parameter
$\sigma_{R}$ remains invariant until the particle passes the minimum distance
between the trajectory and the meteoroid center, $R_{0}$. After this the
negative sign of $\sigma_{R}$ switches to the positive one, i.e., the incoming
particle becomes outgoing. At a given location characterized by $R$, $\theta$
with the given velocity-space parameters $V$, $R_{0}$, $\Phi$, the entire
velocity distribution is composed of two distinct partial$\ $distributions:
one for the incoming particles and the other for the outgoing particles. They
correspond to particles arriving from the two opposite directions. We will
distinguish these two distributions by the corresponding subscripts,
$f_{\sigma_{R}}^{\left(  2\right)  }=f_{\pm}^{\left(  2\right)  }$. Note that
the functions $f_{+}^{\left(  2\right)  }$ and $f_{-}^{\left(  2\right)  }$
can be vastly different. For example, the primary meteoric particles are
exclusively outgoing, so that $f_{-}^{\left(  1\right)  }=0$. Any incoming
particles appear only due to collisions.

The characteristics of equation~(\ref{start}) are straight-line particle
trajectories characterized by $V$, $R_{0}$, $\Phi$ and arriving at a given
location $R$, $\theta$ from the direction determined by $\sigma_{R}$. To
distinguish the fixed coordinate $R$ from flowing coordinates along a given
trajectory we will denote the latter by $R^{\prime}$ with corresponding
$V_{R^{\prime}}$ and $\sigma_{R^{\prime}}$. As discussed in
section~\ref{Collisions of individual primary particles with the atmosphere},
for almost the entire secondary-particle population $\mu=\cos\vartheta$ is
positive, so that all relevant particle trajectories start in front of the
descending meteoroid sufficiently far from it, where there are virtually no
sheath particles.

We can write the formal solution of equation~(\ref{start}) as
\begin{linenomath*}
\begin{equation}
f_{\sigma_{R}}^{\left(  2\right)  }\approx\frac{1}{V}\ \delta\left(
V-\frac{2\mu m_{\beta}U}{m+m_{\beta}}\right)  \int_{\infty}^{R}\frac{K\left(
\mu,R^{\prime}\right)  R^{\prime}dR^{\prime}}{\sigma_{R^{\prime}}%
\sqrt{(R^{\prime})^{2}-R_{0}^{2}}}. \label{f^(2)_approximate}%
\end{equation}
\end{linenomath*}
where the lower integration limit denotes an infinitely far distance in front
of the descending meteoroid. This schematic expression, however, needs a
refinement associated with the fact that the spherical radius $R^{\prime}$ is
a non-monotonic function of the particle position along its trajectory. For
the incoming particles with $\sigma_{R}=-1$, the radial distance $R^{\prime}$
is monotonically decreasing, so that along the entire particle trajectory we
have $\sigma_{R^{\prime}}=-1$. In contrast, for the outgoing particles
($\sigma_{R}=+1$) arriving at a given location $R$, the radial distance
$R^{\prime}$ first decreases ($\sigma_{R^{\prime}}=-1$) from $\infty$ down to
the minimum distance $R^{\prime}=R_{0}$ and then starts increasing again
($\sigma_{R^{\prime}}=+1$) until it reaches $R^{\prime}=R$. As a result, the
integral in equation~(\ref{f^(2)_approximate}) leads to a piecewise
expression:%
\begin{linenomath*}
\begin{equation}
\int_{\infty}^{R}\frac{K\left(  \mu,R^{\prime}\right)  R^{\prime}dR^{\prime}%
}{\sigma_{R^{\prime}}\sqrt{(R^{\prime})^{2}-R_{0}^{2}}}=\left\{  \!\!%
\begin{array}
[c]{ccc}%
\int_{R}^{\infty}\frac{K\left(  \mu,R^{\prime}\right)  R^{\prime}dR^{\prime}%
}{\sqrt{\left(  R^{\prime}\right)  ^{2}-R_{0}^{2}}} & \text{if} & \sigma
_{R}=-1,\\
&  & \\
\int_{R_{0}}^{R}\frac{K\left(  \mu,R^{\prime}\right)  R^{\prime}dR^{\prime}%
}{\sqrt{\left(  R^{\prime}\right)  ^{2}-R_{0}^{2}}}\ {\normalsize +}%
\int_{R_{0}}^{\infty}\frac{K\left(  \mu,R^{\prime}\right)  R^{\prime
}dR^{\prime}}{\sqrt{\left(  R^{\prime}\right)  ^{2}-R_{0}^{2}}} & \text{if} &
\sigma_{R}=+1.
\end{array}
\right.  \label{dz'/dR'}%
\end{equation}
\end{linenomath*}

\begin{figure}[h]
\centering
\includegraphics[width=30pc]
{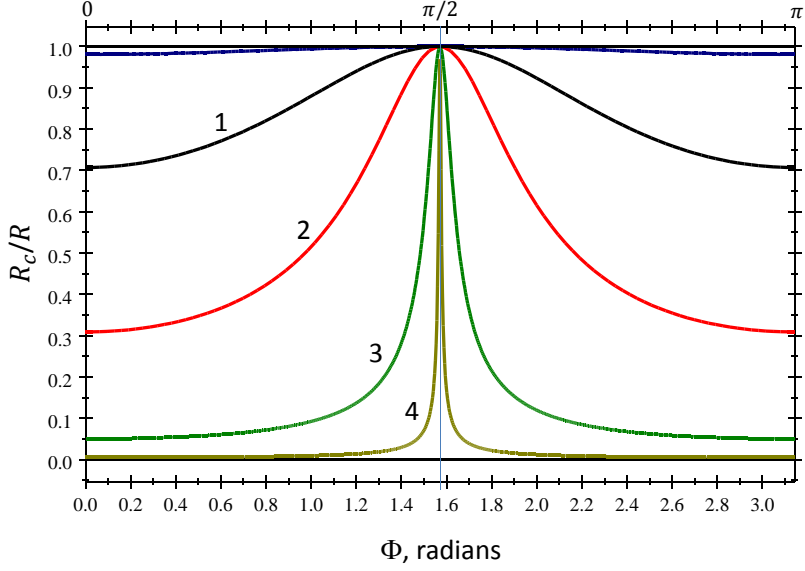}
\caption{The function $R_{c}(\Phi)/R$ for several values of $\theta$: curve~1
for $\theta=\pi/4$, curve~2 for $\theta=3\pi/5$, curve~3 for $\theta
=\pi/2-0.05\approx1.5208$, curve~4 for $\theta=\pi/2-0.005\approx1.5658$.}%
\label{Fig:R_c}
\end{figure}

Another important issue is that the secondary particles have only positive
values of $\mu=\cos\vartheta=(\vec{V}\cdot\vec{U})/(VU)$, as described in
section~\ref{Collisions of individual primary particles with the atmosphere};
otherwise $f_{\sigma_{R}}^{\left(  2\right)  }=0$. The invariant parameter
$\mu$ can be expressed in terms of the spherical coordinates, $R$, $\theta$,
and velocity-space variables, $R_{0}$, $\Phi$, $\sigma_{R}$, as
\begin{linenomath*}
\begin{align}
&  \mu\equiv\cos\vartheta=\cos\Theta\cos\theta+\sin\Theta\sin\theta\cos
\Phi\nonumber\\
&  =\sigma_{R}\sqrt{1-\frac{R_{0}^{2}}{R^{2}}}\ \cos\theta+\frac{R_{0}%
\sin\theta}{R}\ \cos\Phi. \label{mumuha}%
\end{align}
\end{linenomath*}
The boundary $\mu=0$ corresponds to the critical value of $R_{0}=R_{c}(\Phi)$,
\begin{linenomath*}
\begin{equation}
R_{c}(\Phi)=\frac{R\left\vert \cos\theta\right\vert }{\sqrt{\cos^{2}%
\theta+\sin^{2}\theta\cos^{2}\Phi}}=\frac{R\left\vert \cos\theta\right\vert
}{\sqrt{1-\sin^{2}\theta\sin^{2}\Phi}}. \label{R_c(Phi)}%
\end{equation}
\end{linenomath*}
Figure~\ref{Fig:R_c} shows $R_{c}(\Phi)/R$ for a few values of $\theta$. As
$\theta\rightarrow0,\pi$ the entire function $R_{c}(\Phi)$ approaches $R$. As
$\theta\rightarrow\pi/2$ the function approaches $0$ everywhere except narrow
spikes near $\Phi=\pm\pi/2$ where $R_{c}=R$. Behind the descending meteoroid,
where $\cos\theta>0$, we have $f_{\sigma_{R}}^{\left(  2\right)  }=0$ for
$\sigma_{R}=+1$ within $0<R_{0}<R_{c}(\Phi)$ and for $\sigma_{R}=-1$ within
$R_{c}(\Phi)<R_{0}<R$. In front of the meteoroid, where $\cos\theta<0$, on the
contrary, $f_{\sigma_{R}}^{\left(  2\right)  }=0$ for $\sigma_{R}=+1$ within
$R_{c}(\Phi)<R_{0}<R$ and for $\sigma_{R}=-1$ within $0<R_{0}<R_{c}(\Phi)$.%

As a result, given the particle coordinates in real space, $R$, $\theta$, with
the velocity parameters, $R_{0}=R\sin\Theta$, $\Phi$, $\sigma_{R}$, and using
the explicit expression for $K\left(  \mu,R^{\prime}\right)  $ given by
equation~(\ref{K(mu,R)}), we obtain:
\begin{linenomath*}
\begin{align}
&  f_{\sigma_{R}}^{(2)}=L_{\sigma_{R}}\ \delta\left(  V-\frac{2m_{\beta}\mu
U}{m+m_{\beta}}\right)  ,\nonumber\\
L_{\sigma_{R}}  &  =\frac{G(U,1-2\mu^{2})n_{0}n_{\mathrm{Atm}}}{\sqrt{3}\ \mu
U^{2}}\left(  1+\frac{m}{m_{\beta}}\right)  ^{3}I(R,R_{0}) \label{f^(2)_via_I}%
\end{align}
\end{linenomath*}
if, concurrently,%
\begin{linenomath*}
\begin{subequations}
\begin{align}
&  \sigma_{R}=\mathrm{sgn}(\cos\theta)\qquad\text{and}\qquad0<R_{0}<R_{c}%
(\Phi)\label{sigma=sgn}\\
\text{or}  & \nonumber\\
&  \sigma_{R}=-\mathrm{sgn}(\cos\theta)\qquad\text{and}\qquad R_{c}%
(\Phi)<R_{0}<R. \label{sigma=-sgn}%
\end{align}
\end{subequations}
\end{linenomath*}
Otherwise (i.e., if $\mu<0$), $f_{\sigma_{R}}^{\left(  2\right)  }=0$. In
accord with equation~(\ref{dz'/dR'}), the multiplier $I(R,R_{0})$ in
(\ref{f^(2)_via_I}), has a piecewise definition
\begin{linenomath*}
\begin{equation}
I(R,R_{0})=\left\{
\begin{array}
[c]{ccc}%
J_{R}^{\infty} & \text{for} & \sigma_{R}=-1\\
&  & \\
J_{R_{0}}^{\infty}\mathrm{H}\left(  R_{0}-r_{\mathrm{M}}\right)
+J_{\max\left(  R_{0},r_{\mathrm{M}}\right)  }^{R} & \text{for} & \sigma
_{R}=+1,
\end{array}
\right.  \label{III}%
\end{equation}
\end{linenomath*}
where the integral $J_{a}^{b}$, as a function of its integration limits,
$b>a\geq R_{0}$, is defined by
\begin{linenomath*}
\begin{align}
&  J_{a}^{b}\equiv2\int_{a}^{b}\left(  1-\sqrt{1-\frac{r_{\mathrm{M}}^{2}%
}{\left(  R^{\prime}\right)  ^{2}}}\right)  \left[  1+\left(  \frac{R^{\prime
}}{\lambda^{(1)}}\right)  ^{2/3}\right] \nonumber\\
&  \times\exp\left[  -\ \frac{3}{2}\left(  \frac{R^{\prime}}{\lambda^{(1)}%
}\right)  ^{2/3}\right]  \frac{R^{\prime}dR^{\prime}}{\sqrt{\left(  R^{\prime
}\right)  ^{2}-R_{0}^{2}}}\ . \label{J_a^b}%
\end{align}
\end{linenomath*}

The first line in the RHS of equation~(\ref{III}) describes incoming particles
that arrive at a given location $\vec{R}$ after being scattered or ionized
within the segment of a given straight-line trajectory located between an
infinitely large distance (in front of the descending meteoroid) and $\vec{R}%
$. The second line in the RHS of equation~(\ref{III}) describes outgoing
particles that arrive along a different line segment from the opposite
direction. The first term, $J_{R_{0}}^{\infty}\mathrm{H}\left(  R_{0}%
-r_{\mathrm{M}}\right)  $, describes secondary particles originated in the
beginning part of the straight-line trajectory, from an infinitely large
distance down to the minimum distance $R_{0}$. In this part of the trajectory
the particles were incoming. The second term, $J_{R_{0}}^{R}$, describes the
outgoing particles originated within the remaining part of this trajectory
$R$, as illustrated by Figure~\ref{Fig:Scheme}. The step-function $\mathrm{H}%
\left(  R_{0}-r_{\mathrm{M}}\right)  $ and $\max\left(  R_{0},r_{\mathrm{M}%
}\right)  $ there are associated with particle trajectories that may cross the
meteoroid surface, $R_{0}<r_{\mathrm{M}}$. Such trajectories always exist,
regardless of how far from the meteoroid is the given location $\vec{R}$. For
the trajectories with $R_{0}<r_{\mathrm{M}}$, the meteoroid surface shields
all outgoing particles arriving at $\vec{R}$ from the opposite side of the
meteoroid. This shielding will inevitably lead to a dip in the velocity
distribution of the outgoing particles, $f_{+}^{(2)}$, with a discontinuity at
$R_{0}=r_{\mathrm{M}}$.

The velocity distribution of secondary particles that originated in close
proximity to the meteoroid is sensitive to the actual meteoroid shape. The
latter can be far from spherical, and even more so, the actual boundary
conditions on the meteoroid surface may include inelastic reflections of the
impinging particles from chaotically distributed surface irregularities. In
this case, the anticipated dip in $f_{+}^{(2)}$ might be, at least partially,
filled with such reflected particles. Since we do not have a detailed
knowledge of these conditions, we will stick to the simplest case of the
ideally spherical meteoroid with the fully absorbing surface. We will also
ignore local disturbances of the neutral atmosphere that are caused by the
falling meteoroid itself.

In equation (\ref{J_a^b}), the integral $J_{a}^{b}$ is analytically
intractable and needs approximations. Presenting $J_{\max\left(
R_{0},r_{\mathrm{M}}\right)  }^{R}$ as the difference of two well-convergent
improper integrals, $J_{\max\left(  R_{0},r_{\mathrm{M}}\right)  }^{R}%
=J_{\max\left(  R_{0},r_{\mathrm{M}}\right)  }^{\infty}-J_{R}^{\infty}$, we
restrict our analysis to $J_{a}^{\infty}$, where the lower integration limit,
$a$, equals $R$, $R_{0}$, or $r_{\mathrm{M}}$. These integrals allow
analytical approximations due to the fact that the mean free path of the
primary particles, $\lambda^{(1)}$, is many orders of magnitude larger than
the meteoroid size, $r_{\mathrm{M}}$. This allows separating the entire domain
of parameters $R$ and $R_{0}$ into overlapped sub-domains where the integral
(\ref{J_a^b}) becomes simpler. On the one hand, for relatively large $a\gg
r_{\mathrm{M}}$, including $a\gtrsim\lambda^{(1)}$, the first factor in the
integrand of (\ref{J_a^b}) allows the Taylor expansion $2(1-\sqrt
{1-r_{\mathrm{M}}^{2}/(R^{\prime})^{2}}\ )\approx r_{\mathrm{M}}%
^{2}/(R^{\prime})^{2}$. This expansion works well even under a much weaker
condition of $R^{\prime}\gtrsim3r_{\mathrm{M}}$. On the other hand, for
relatively short distances from the meteoroid center, $R_{0}\leq R\ll
\lambda^{(1)}$, including $R_{0}\lesssim r_{\mathrm{M}}$, in the entire range
of $R^{\prime}\gtrsim\lambda^{(1)}$ the integrand of (\ref{J_a^b}) is small
and makes no tangible contribution to the total integral value, while for
$R^{\prime}\ll\lambda^{(1)}$ one can neglect the terms involving $(R^{\prime
}/\lambda^{(1)})^{2/3}$. This allows the remaining integral to be expressed in
terms of the elliptic integrals. Below we calculate $J_{a}^{\infty}$
separately for each sub-domain.

\paragraph{Near zone: $r_{\mathrm{M}}\leq R\ll\lambda^{(1)}$.}

The corresponding integrals are calculated in the appendix in terms of the
elliptic integrals. This calculation yields
\begin{linenomath*}
\begin{equation}
J_{R}^{\infty}\approx\left\{
\begin{array}
[c]{ccc}%
2r_{\mathrm{M}}\mathrm{E}\left(  \frac{r_{\mathrm{M}}}{R};\frac{R_{0}%
}{r_{\mathrm{M}}}\right)  -2\sqrt{R^{2}-R_{0}^{2}}\left(  1-\sqrt
{1-\frac{r_{\mathrm{M}}^{2}}{R^{2}}}\right)  & \text{if} & R_{0}\leq
r_{\mathrm{M}}\\
&  & \\%
\begin{array}
[c]{c}%
2R_{0}\left[  \mathrm{E}\left(  \frac{R_{0}}{R};\frac{r_{\mathrm{M}}}{R_{0}%
}\right)  -\left(  1-\frac{r_{\mathrm{M}}^{2}}{R_{0}^{2}}\right)
\ \mathrm{F}\left(  \frac{R_{0}}{R};\frac{r_{\mathrm{M}}}{R_{0}}\right)
\right] \\
\\
-\ 2\left(  1-\sqrt{1-\frac{r_{\mathrm{M}}^{2}}{R^{2}}}\right)  \sqrt
{R^{2}-R_{0}^{2}}%
\end{array}
& \text{if} & R_{0}\geq r_{\mathrm{M}},
\end{array}
\right.  \label{I}%
\end{equation}
\end{linenomath*}
and
\begin{linenomath*}
\begin{equation}
J_{\max\left(  R_{0},r_{\mathrm{M}}\right)  }^{\infty}\approx\left\{
\begin{array}
[c]{ccc}%
2r_{\mathrm{M}}\mathrm{E}\left(  \frac{R_{0}}{r_{\mathrm{M}}}\right)
-2\sqrt{r_{\mathrm{M}}^{2}-R_{0}^{2}} & \text{if} & R_{0}\leq r_{\mathrm{M}}\\
&  & \\
2R_{0}\left[  \mathrm{E}\left(  \frac{r_{\mathrm{M}}}{R_{0}}\right)  -\left(
1-\frac{r_{\mathrm{M}}^{2}}{R_{0}^{2}}\right)  \mathrm{K}\left(
\frac{r_{\mathrm{M}}}{R_{0}}\right)  \right]  & \text{if} & R_{0}\geq
r_{\mathrm{M}},
\end{array}
\right.  . \label{I(R_0,R_0)}%
\end{equation}
\end{linenomath*}

\paragraph{Intermediate sub-domain: $3r_{\mathrm{M}}\lesssim R_{0}\leq
R\ll\lambda^{(1)}$.}

These conditions enable the easiest calculation of $J_{a}^{\infty}$ and, at
the same time, they cover a rather broad parameter sub-domain. Using
simultaneously both Taylor expansions described above and restricting them to
the highest-order terms, we arrive at fairly simple expressions:
\begin{linenomath*}
\begin{equation}
J_{R}^{\infty}\approx\frac{r_{\mathrm{M}}^{2}}{R_{0}}\ \arcsin\frac{R_{0}}%
{R},\qquad J_{\max\left(  R_{0},r_{\mathrm{M}}\right)  }^{\infty}=J_{R_{0}%
}^{\infty}\approx\frac{\pi r_{\mathrm{M}}^{2}}{2R_{0}}. \label{J_a^b_arcsin}%
\end{equation}
\end{linenomath*}

\paragraph{Long-distance/large-$R_{0}$ zone: $R\geq R_{0}\gg r_{\mathrm{M}}$.}

In this sub-domain, we should keep all the factors with $(R_{0}/\lambda
^{(1)})^{2/3}$, but can use the Taylor expansion $2(1-\sqrt{1-r_{\mathrm{M}%
}^{2}/(R^{\prime})^{2}})\approx r_{\mathrm{M}}^{2}/(R^{\prime})^{2}$. Then we
temporarily recast $J_{a}^{\infty}$ with $a\gg r_{\mathrm{M}}$ as
\begin{linenomath*}
\begin{equation}
J_{a}^{\infty}=\frac{r_{\mathrm{M}}^{2}I_{\alpha}^{\infty}}{R_{0}},\qquad
I_{\alpha}^{\infty}\left(  \Lambda\right)  =\int_{\alpha}^{\infty}\left(
\Lambda+\frac{3}{2y}\right)  \frac{e^{-\Lambda y}dy}{\sqrt{y^{3}-1}},
\label{I_R_0_reduced}%
\end{equation}
\end{linenomath*}
where
\begin{linenomath*}
\begin{equation}
y\equiv\left(  \frac{R^{\prime}}{R_{0}}\right)  ^{2/3},\qquad\alpha
\equiv\left(  \frac{a}{R_{0}}\right)  ^{2/3},\qquad\Lambda\equiv\frac{3}%
{2}\left(  \frac{R_{0}}{\lambda_{T}^{(1)}}\right)  ^{2/3}. \label{y_reduced}%
\end{equation}
\end{linenomath*}
In the limit of $\Lambda\alpha\ll1$ corresponding to $a\gg r_{\mathrm{M}}$, by
changing variable $y=(1+z^{2})^{1/3}$ one can easily verify that
equation~(\ref{I_R_0_reduced}) yields (\ref{J_a^b_arcsin}).

Calculating the integral $I_{\alpha}^{\infty}\left(  \Lambda\right)  $ in
limiting cases and interpolating between those, we construct the following
approximation,
\begin{linenomath*}
\begin{equation}
I_{\alpha}^{\infty}\left(  \Lambda\right)  \approx\frac{e^{-\Lambda\alpha}%
}{\sqrt{\alpha^{3}-1+\left[  \pi\left(  \frac{\Lambda}{3}+\frac{\pi}%
{4}\right)  \right]  ^{-1}}}. \label{J_lambda>>1}%
\end{equation}
\end{linenomath*}
Direct numeric calculations show that the maximum discrepancy between the
exact value of the integral and that given by equation~(\ref{J_lambda>>1}) is
near $\alpha\simeq1.06$, where it reaches $\simeq12\%$; in most other
occasions this discrepancy is much smaller. In the original notations,
equation~(\ref{J_lambda>>1}) yields
\begin{linenomath*}
\begin{equation}
J_{R}^{\infty}\approx\frac{r_{\mathrm{M}}^{2}\sqrt{1+\frac{2}{\pi}\left(
\frac{R_{0}}{\lambda_{T}^{(1)}}\right)  ^{2/3}}\exp\left[  -\ \frac{3}%
{2}\left(  \frac{R}{\lambda_{T}^{(1)}}\right)  ^{2/3}\right]  }{\sqrt{\left[
1+\frac{2}{\pi}\left(  \frac{R_{0}}{\lambda_{T}^{(1)}}\right)  ^{2/3}\right]
\left(  R^{2}-R_{0}^{2}\right)  +\frac{4R_{0}^{2}}{\pi^{2}}}}.
\label{J_a^infty_interpolation}%
\end{equation}
\end{linenomath*}
Setting in equation~(\ref{J_a^infty_interpolation}) $R=R_{0}$, we obtain
\begin{linenomath*}
\begin{equation}
J_{\max(R_{0},r_{\mathrm{M}})}^{\infty}=J_{R_{0}}^{\infty}\approx\frac{\pi
r_{\mathrm{M}}^{2}}{2R_{0}}\sqrt{1+\frac{2}{\pi}\left(  \frac{R_{0}}%
{\lambda_{T}^{(1)}}\right)  ^{2/3}}\exp\left[  -\ \frac{3}{2}\left(
\frac{R_{0}}{\lambda_{T}^{(1)}}\right)  ^{2/3}\right]  .
\label{J_R_0_interpolation}
\end{equation}
\end{linenomath*}
For $R_{0}\ll R\ll\lambda_{T}^{(1)}$, these expressions agree with the
intermediate asymptotics given by equation~(\ref{J_a^b_arcsin}). These
equations give the expressions for $J_{R}^{\infty}$ and $J_{R_{0}}^{\infty
}\mathrm{H}\left(  R_{0}-r_{\mathrm{M}}\right)  +J_{\max\left(  R_{0}%
,r_{\mathrm{M}}\right)  }^{R}$, where $J_{\max\left(  R_{0},r_{\mathrm{M}%
}\right)  }^{R}=J_{\max\left(  R_{0},r_{\mathrm{M}}\right)  }^{\infty}%
-J_{R}^{\infty}$, to be substituted first in equation~(\ref{III}) and then to
(\ref{f^(2)_via_I}) for calculating $f_{\sigma_{R}}^{(2)}$.

All the above relations are expressed in terms of the invariant variables
$R_{0}$, $\mu,$ and $V$. For applications, it is more convenient to express
$R_{0}$ in terms of the polar angle in velocity space, $\Theta$, around the
local radius-vector $\vec{R}$ direction, $R_{0}=R\sin\Theta$. One might also
need to express $\mu$ in terms of $\Theta$ and the corresponding axial angle,
$\Phi$ (measured from the common $\vec{R}$-$\vec{U}$ plane), $\mu\equiv
\cos\vartheta=\cos\Theta\cos\theta+\sin\Theta\sin\theta\cos\Phi$, where
$\theta$ is the polar angle in real space measured from the direction of
$\vec{U}$.

Figures~\ref{Fig:I(R,RsinTheta)}, \ref{Fig:Ang3D_1}, and \ref{Fig:Ang3D_2}
illustrate $f_{\sigma_{R}}^{(2)}$ given by equation (\ref{f^(2)_via_I}). In
the entire 3-D velocity space, the distribution function has a shell-like
structure depicted by Figure~\ref{Fig:Shell} and approximated by the $\delta
$-function. Below we show the angular dependence of the factor $L_{\sigma_{R}%
}$ preceding the $\delta$-function and defined by equation (\ref{f^(2)_via_I}%
).
\begin{figure}[h]
\centering
\includegraphics[width=30pc]
{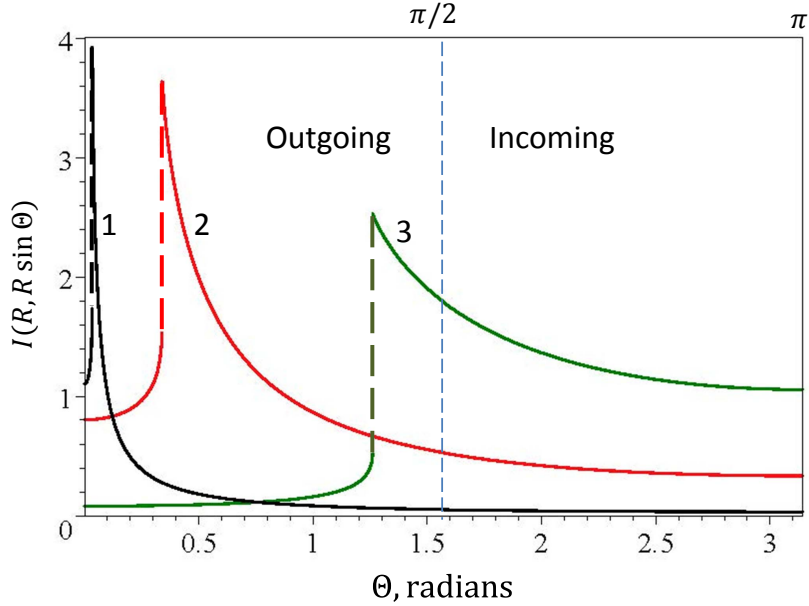}
\caption{The factor $I(R,R_{0})$ with $R_{0}=R\sin\Theta$ for several
distances from the meteoroid surface: curve~1 for $R=30r_{\mathrm{M}}$,
curve~2 for $R=3r_{\mathrm{M}}$, curve 3 for $R=1.05r_{\mathrm{M}}$. The
dashed line segments of each curve indicate trajectories tangential to the
meteoroid surface, $R\sin\Theta=r_{\mathrm{M}}$. For $\Theta<\Theta
_{c}=\arcsin(r_{\mathrm{M}}/R)$, a significant fraction of particles moving
along the corresponding straight-line trajectory are shielded by the meteoroid
and cannot reach the destination point. This shielding produces the pronounced
dips for all $\Theta<\Theta_{c}$. Notice the smooth transition between the
outgoing and incoming particles at $\Theta=\pi/2$.}
\label{Fig:I(R,RsinTheta)}
\end{figure}

We start from the function $I(R,R_{0})=I(R,R\sin\Theta)$ which is the major
$\Theta$-dependent multiplier in the expression for $L_{\sigma_{R}}%
(\Theta,\Phi)$. Figure~\ref{Fig:I(R,RsinTheta)} shows $I(R,R\sin\Theta)$ as a
function of $\Theta$ for several radial distances $R$. Each curve combines $I$
for the outgoing particles, $\sigma_{R}=+1$, with that for the incoming
particles, $\sigma_{R}=-1$. For the outgoing ($0\leq\Theta\leq\pi/2$) and
incoming ($\pi/2\leq\Theta\leq\pi$) particles, equation (\ref{III}) gives
different analytic expressions, but they match at $\Theta=\pi/2$ smoothly. At
$\Theta=\Theta_{\mathrm{cr}}\equiv\arcsin(r_{\mathrm{M}}/R)$, the distribution
function undergoes a discontinuity corresponding to the boundary between the
unhindered particle trajectories and those with particles shielded by the
meteoroid, $R_{0}=r_{\mathrm{M}}$, as we discussed above. The function
$I(R,R\sin\Theta)$ reaches its maximum at $\Theta=\Theta_{\mathrm{cr}}$. For
$R\gg r_{\mathrm{M}}$, the outgoing particles form a spiky angular
distribution at small $\Theta\lesssim r_{\mathrm{M}}/R$ corresponding to
almost radially propagating particles, $\vartheta\approx\theta$. As $R$
becomes large, the narrower this angular spike becomes.

Figure~7 corresponds to distances $R\ll\lambda_{T}^{(1)}$, but for
$R\gtrsim\lambda_{T}^{(1)}$ the angular distribution of $I(R,R\sin\Theta)$ is
qualitatively the same, although resolving the narrow spike on the
corresponding diagram would be hard. The spike about the direction of the
radius-vector $\vec{R}$ occurs because the source density for the secondary
particles is proportional to the density of the primary particles,
$n^{(1)}\propto1/(R^{\prime})^{2}$, and hence is largest near the meteoroid
surface.
\begin{figure}[h]
\centering
\includegraphics[width=30pc]
{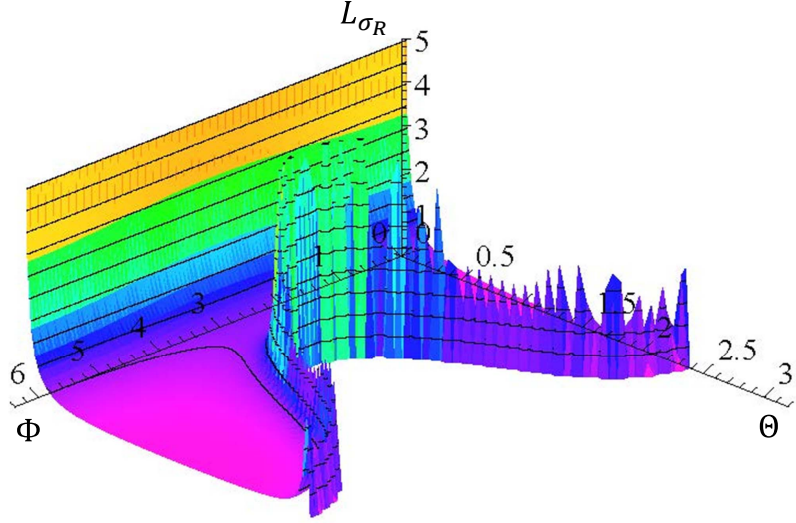}
\caption{The factor $L_{\sigma_{R}}$ defined by equation (\ref{f^(2)_via_I}),
as a function of local polar ($\Theta$) and axial ($\Phi$) angles in the
velocity space, for a location partially behind the descending meteoroid,
$\theta=\pi/4$ (the 3-D surface plot is cut off vertically).}
\label{Fig:Ang3D_1}
\end{figure}
\begin{figure}[h]
\centering
\includegraphics[width=30pc]
{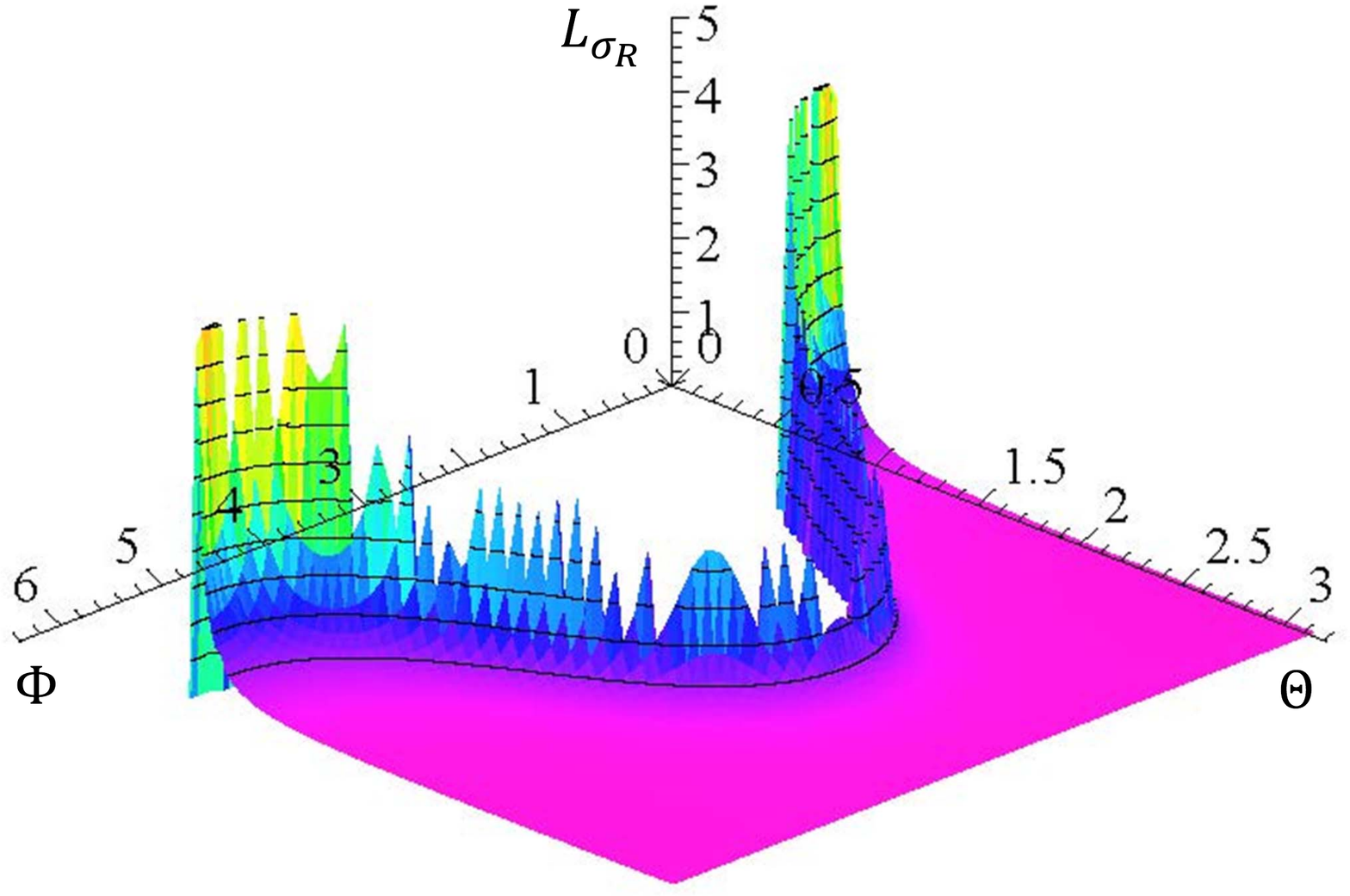}
\caption{The factor $L_{\sigma_{R}}$ defined by equation (\ref{f^(2)_via_I}),
as a function of local polar ($\Theta$) and axial ($\Phi$) angles in the
velocity space, for a location partially in front of the descending meteoroid,
$\theta=3\pi/4$ (the 3-D surface plot is cut off vertically).}
\label{Fig:Ang3D_2}
\end{figure}

Figures~\ref{Fig:Ang3D_1} and \ref{Fig:Ang3D_2} show two examples of the full
factor $L_{\sigma_{R}}(\Theta,\Phi)$ plotted as a 3-D surface over the entire
$\Theta$,$\Phi$-domain: $0\leq\Theta\leq\pi$, $0\leq\Phi\leq2\pi$. Each plot
combines $L_{\sigma_{R}}(\Theta,\Phi)$ for the outgoing particles, $L_{+}$
($0\leq\Theta\leq\pi/2$) with that for the incoming particles, $L_{-}$
($\pi/2\leq\Theta\leq\pi$). Following $I(R,R\sin\Theta)$, the two surfaces of
$L_{\pm}(\Theta,\Phi)$ match smoothly at $\Theta=\pi/2$. The two different 3-D
plots correspond to the same radial distance ($R=30r_{\mathrm{M}}$), but
different values of the polar angle in real space, $\theta$. One corresponds
to an axially symmetric location behind the descending meteoroid ($\theta
=\pi/4$), while the other to a similar location in front of it ($\theta
=3\pi/4$). For simplicity, we assumed an isotropic differential cross-section,
$G(U,1-2\mu^{2})=G(U)$. The non-zero values of $L_{\sigma_{R}}$ occupy a part
of the entire $\Theta$,$\Phi$-domain bounded by $\Theta=0$, $0\leq\Phi\leq
2\pi$ and the $\Theta$,$\Phi$-curve determined by $\mu=\cos\Theta\cos
\theta+\sin\Theta\sin\theta\cos\Phi=0$ (this curve corresponds to $R_{0}%
=R_{c}(\Phi)$ in Figure~7). Near these two boundaries, the plotted surface has
two pronounced `ridges' (for better visualization, they are cut off
vertically). The `ridge' near $\Theta=0$ is formed by the spiky maximum of
$I(R,R\sin\Theta)$ for the outgoing particles, $L_{+}$, as discussed above.
The `ridge' near $\mu=0$ is associated with the singular factor $\mu$ in the
denominator of (\ref{f^(2)_via_I}). As we discuss in the companion paper, this
singularity plays no role for the total plasma density or flux because the
corresponding integrals of $f_{\sigma_{R}}^{(2)}$ include weighting factors
that not only suppress the singularity, but make its contribution negligible.

The `ridge' near $\Theta=0$ is different. Its tallest part with $\Theta
\lesssim3r_{\mathrm{M}}/R$ corresponding to $R_{0}\lesssim3r_{\mathrm{M}}$
comes from trajectories passing through a small volume near the meteoroid
surface where the density of the primary particles increases $\propto
1/(R^{\prime})^{2}$. This part of the distribution function is described by
more complicated equations~(\ref{I}) and (\ref{I(R_0,R_0)}). However, for most
locations with $R\gg r_{\mathrm{M}}$, particles with $\Theta\lesssim
3r_{\mathrm{M}}/R$ cannot make significant contributions to the total particle
density and flux because the contribution of secondary particles originated
beyond the near-meteoroid volume are $\sim\lambda_{T}^{(1)}/r_{\mathrm{M}}$
times larger. These particles form a `pedestal' in the $\Theta$-distribution
of secondary particles which at large distances $R$ is localized,
$\Theta\lesssim\lambda_{T}^{(1)}/R$, but much broader than the central spike
of $\Theta\lesssim3r_{\mathrm{M}}/R$. The fact that in spite of the much
higher density of primary particles directly near the meteoroid, $R^{\prime
}\lesssim3r_{\mathrm{M}}$, the majority of secondary particles exists within a
large volume $r_{\mathrm{M}}\ll R^{\prime}\lesssim\lambda_{T}^{(1)}$. The
scattering or ionization of the primary particles is described by the
exponential factor $\propto\exp[-(3/2)(R^{\prime}/\lambda_{T}^{(1)})^{2/3}]$
in equation (\ref{J_a^b}). This exponential loss of $\ll$primary particles
represents the entire source for the secondary particles. Hence the volume
with $R^{\prime}\ll\lambda_{T}^{(1)}$ with almost no exponential loss cannot
be the dominant source of secondary particles.

\section{Summary and conclusions\label{Summary and conclusions}}

We have developed a first-principle collisional kinetic model of plasma and
neutral sheath formed around a fast-descending small meteoroid when it passes
the altitude range of 90-120 km. In this range, sensitive radars detect
atmospheric effects of the meteoroid passage and we will use this theory to
more accurately interpret the radar head echo \citep{Close:NewMethod2005,Campbell-Brown:Meteoroid2007}.
The analytic theory of this paper
describes the spatial structure and velocity distributions of heavy particles:
ions and neutral particles of the meteoric origin, while electrons are assumed
to roughly follow the Boltzmann distribution.

The velocity distribution of the neutral `primary' particles, $f^{(1)}$, is
given by equation~(\ref{f^(0)_general}), but the central topic of this paper
is finding the velocity distributions of the `secondary' particles, $f^{(2)}$,
where we have also included all higher-order subgroups, $f^{(j)}$ with $j>2$.
These distributions are described by equations (\ref{f^(2)_via_I}) and
(\ref{III}). If $3r_{\mathrm{M}}\lesssim R_{0}\leq R\ll\lambda_{T}^{(1)}$ then
$J_{a}^{\infty}$ reduce to simple equation (\ref{J_a^b_arcsin}). An important
feature of the `secondary'-particle distribution functions $f_{\sigma_{R}%
}^{(2)}$ ($\sigma_{R}=\pm1$) is their shell-like distribution in velocity
space, approximately described in equation~(\ref{f^(2)_via_I}) by the $\delta
$-function factor and illustrated by Figure~\ref{Fig:Shell}. Such anisotropic
and non-monotonic velocity distributions are potentially unstable and may give
rise to detectable plasma turbulence. However, analysis of kinetic plasma
instabilities in the near-meteoroid plasma may require a more accurate
description of electrons.

In the companion paper, we apply this kinetic theory to calculate a spatial
structure of the near-meteoroid plasma. This will allow accurate modeling of
radar head echoes. Our future work will include a more detailed and specific
theoretical analysis, computer simulations, and discussion of implications for
real meteors and comparisons with radar and other observations.

\acknowledgments
This work was supported by NSF Grant AGS-1244842.

\newpage

\appendix

\section{Derivation of kinetic equation~(\ref{D_t(f)})\label{Derivation of kinetic Eq}}

We start with the collisional kinetic equation for $f(\vec{V},\vec{r})$ in a
general probabilistic form \citep{Lifshitz:Physical1981},
\begin{linenomath*}
\begin{align}
&  \frac{\partial}{\partial t}+\vec{V}\cdot\nabla+\frac{\vec{F}}{m}\cdot
\frac{\partial f}{\partial\vec{V}}=S_{\mathrm{arr}}-S_{\mathrm{dep}%
},\nonumber\\
&  S_{\mathrm{arr}}=\sum_{\beta}\int f^{\prime}f_{\beta}^{\prime}W(\vec
{V}^{\prime}\rightarrow\vec{V})\delta\left(  E^{\prime}+E_{\beta}^{\prime
}-\Delta E_{\mathrm{in}}-E-E_{\beta}\right) \nonumber\\
&  \times\delta\left(  m\vec{V}^{\prime}+m_{\beta}\vec{V}_{\beta}^{\prime
}-m\vec{V}-m_{\beta}\vec{V}_{\beta}\right)  d^{3}V_{\beta}d^{3}V_{\beta
}^{\prime}d^{3}V^{\prime},\label{D_t(f)_init}\\
&  S_{\mathrm{dep}}=f\sum_{\beta}\int f_{\beta}W(\vec{V}\rightarrow\vec
{V}^{\prime})\delta\left(  E^{\prime}+E_{\beta}^{\prime}+\Delta E_{\mathrm{in}%
}-E-E_{\beta}\right) \nonumber\\
&  \times\delta\left(  m\vec{V}^{\prime}+m_{\beta}\vec{V}_{\beta}^{\prime
}-m\vec{V}-m_{\beta}\vec{V}_{\beta}\right)  d^{3}V_{\beta}d^{3}V_{\beta
}^{\prime}d^{3}V^{\prime},\nonumber
\end{align}
\end{linenomath*}
where in the LHS $\vec{F}$ describes external forces and the RHS presents the
combined operator of binary collisions. The integral term $S_{\mathrm{arr}}$
describes the collisional \textquotedblleft gain\textquotedblright\ of
particles at a given infinitesimal 6-D phase-space volume $d^{3}Vd^{3}%
V_{\beta}$ around $\vec{V}$ and $\vec{V}_{\beta}$, while $S_{\mathrm{dep}}$
describes the corresponding collisional \textquotedblleft
departure\textquotedblright\ from this volume after one collision act.

The function $W(\vec{V}^{\prime}\rightarrow\vec{V})$ denotes the probability
per unit time that two colliding particles with the velocities $\vec
{V}^{\prime},\vec{V}_{\beta}^{\prime}$ within the elementary phase volumes
$d^{3}V^{\prime}$ and $d^{3}V_{\beta}^{\prime}$ will be collisionally
scattered into particles with the velocities $\vec{V},\vec{V}_{\beta}$ within
the volumes $d^{3}V$ and $d^{3}V_{\beta}$. The arrows in the symbolic
arguments of $W$ indicate the order in which the given collision act takes
place. As in \citet{Huang:Statistical87}, here we explicitly factored out the
$\delta$-functions that express the conservation of the total particle energy
and momentum of all colliding particles during one collision act. Unlike
\citet{Lifshitz:Physical1981,Huang:Statistical87}, however, we assume
here that some collisions can be inelastic, resulting in excitation of
internal molecular/atomic degrees of freedom or ionization; the corresponding
energy losses are denoted by $\Delta E_{\mathrm{in}}$. As explained in
section~\ref{Formulation of the kinetic problem}, for simplicity we consider
only one kind of inelastic collisions with the given discrete energy loss
$\Delta E_{\mathrm{in}}$.

All available classical and quantum-mechanical models of binary collisions
operate with the differential cross-sections, $d\sigma/d\Omega$, rather than
with the probabilities, $W$. Hence we need to express $W$ in terms of
$d\sigma/d\Omega$. It is easier to do for the departure term, $S_{\mathrm{dep}%
}$, because it does not involve the given distribution function $f$ in its
integrand. The expression for $S_{\mathrm{dep}}$ in (\ref{D_t(f)_init}) is
proportional to $f$, so that the remaining integral factor is the
corresponding kinetic collision frequency, $\nu(\vec{V})$: $S_{\mathrm{dep}%
}=\nu f$. In $S_{\mathrm{dep}}$ the velocities $\vec{V}$ and $\vec{V}_{\beta}$
describe colliding particles before their collision, while $\vec{V}^{\prime
},\vec{V}_{\beta}^{\prime}\ $describe the same particles immediately after it.
If this is an ionizing collision then the primed variables describe the newly
born ions.

Reducing $S_{\mathrm{dep}}$ to the more traditional Boltzmann form contains
several steps. First, we integrate $S_{\mathrm{dep}}$ over $d^{3}V_{\beta
}^{\prime}$ with elimination of the momentum $\delta$-function, $\delta
(m\vec{V}^{\prime}+m_{\beta}\vec{V}_{\beta}^{\prime}-m\vec{V}-m_{\beta}\vec
{V}_{\beta})$. This yields the factor $m_{\beta}^{-3}$ and leads to the
conservation of the total momentum, but not yet to the energy conservation.
Further, we pass from the integration over $d^{3}V^{\prime}$ to the
integration over $d^{3}u^{\prime}=\left(  u^{\prime}\right)  ^{2}du^{\prime
}d\Omega_{\mathrm{s}}$ where $u^{\prime}=|\vec{u}^{\prime}|$, $\vec{u}%
^{\prime}=\vec{V}^{\prime}-\vec{V}_{\beta}^{\prime}$, and $d\Omega
_{\mathrm{s}}=\left(  \sin\Theta_{\mathrm{s}}\right)  d\Theta_{\mathrm{s}%
}d\Phi_{\mathrm{s}}=d\Lambda d\Phi_{\mathrm{s}}.$ Integrating over
$du^{\prime}$ we eliminate the remaining $\delta$-function,%
\begin{linenomath*}
\begin{equation}
\int\left(  \cdots\right)  \delta\left(  E^{\prime}+E_{\beta}^{\prime}+\Delta
E_{\mathrm{in}}-E-E_{\beta}\right)  \left(  u^{\prime}\right)  ^{2}du^{\prime
}=\frac{\left(  \cdots\right)  \left(  u^{\prime}\right)  ^{2}}{\partial
(E^{\prime}+E_{\beta}^{\prime})/\partial u^{\prime}},\label{promezh}%
\end{equation}
\end{linenomath*}
and obtain the total energy conservation with inelastic losses,
\begin{linenomath*}
\begin{equation}
E^{\prime}+E_{\beta}^{\prime}=E+E_{\beta}-\Delta E_{\mathrm{in}}%
.\label{energy_conservation}%
\end{equation}
\end{linenomath*}
In equation~(\ref{promezh}) $\partial(E^{\prime}+E_{\beta}^{\prime})/\partial
u^{\prime}$ must be calculated before applying
equation~(\ref{energy_conservation}). Expressing the individual particle
velocities in terms of $\vec{u}$, $\vec{u}^{\prime}$ according to
Eq.~(\ref{VV_COM}), we obtain $d^{3}V^{\prime}=[m/(m+m_{\beta})]^{3}\left(
u^{\prime}\right)  ^{2}du^{\prime}d\Omega$, while equation~(\ref{E+E}) allows
to easily calculate $\partial(E^{\prime}+E_{\beta}^{\prime})/\partial
u^{\prime}$. After having expressed all quantities in the integrand of
$S_{\mathrm{dep}}$ in terms of the relative particle velocities, $V_{\beta}$
becomes involved only in $f_{\beta}$. Then we can easily integrate over
$d^{3}V_{\beta}$ using $\int f_{\beta}d^{3}V_{\beta}=n_{\mathrm{A}}$ and
obtain
\begin{linenomath*}
\begin{equation}
S_{\mathrm{dep}}=\left(  \frac{1}{m+m_{\beta}}\right)  ^{2}\frac
{fn_{\mathrm{A}}}{mm_{\beta}}\int W(\vec{V}\rightarrow\vec{V}^{\prime
})u^{\prime}d\Omega_{\mathrm{s}}.\label{nu_primary}%
\end{equation}
\end{linenomath*}
Now we compare Eq.~(\ref{nu_primary}) with the conventional Boltzmann
expression \citep{Lifshitz:Physical1981},
\begin{linenomath*}
\begin{equation}
S_{\mathrm{dep}}=f\int u\ \frac{d\sigma}{d\Omega}\ f_{\beta}d^{3}V_{\beta
}d\Omega_{\mathrm{s}}=fn_{\mathrm{A}}\iint\frac{d\sigma}{d\Omega}%
\ ud\Omega_{\mathrm{s}}.\label{S_dep_standard}%
\end{equation}
\end{linenomath*}
Switching from $d\sigma(u,\Theta_{s})/d\Omega$ to $G(u,\Lambda)$, as described
above equation~(\ref{u,Lambda,Phi}), and integrating in $d\Omega_{\mathrm{s}%
}=d\Lambda d\Phi_{\mathrm{s}}$ over $\Phi_{\mathrm{s}}$, we obtain the final
form for $S_{\mathrm{dep}}$ with the corresponding collision frequency, $\nu$,
\begin{linenomath*}
\begin{equation}
S_{\mathrm{dep}}=\nu(\vec{V})f,\qquad\nu(\vec{V})=2\pi n_{\mathrm{A}}\int%
_{-1}^{1}uG(u,\Lambda)d\Lambda.\label{nu_final}%
\end{equation}
\end{linenomath*}
Comparing equations~(\ref{nu_primary}) and (\ref{nu_final}), we obtain the
expression for the probability $W(\vec{V}\rightarrow\vec{V}^{\prime})$ in
terms of the differential cross-section,
\begin{linenomath*}
\begin{equation}
W(\vec{V}\rightarrow\vec{V}^{\prime})=mm_{\beta}\left(  m+m_{\beta}\right)
^{2}\left(  \frac{u}{u^{\prime}}\right)  G(u,\Lambda
),\label{W(P->P')_via_d_sigma}%
\end{equation}
\end{linenomath*}
where $u$ and $u^{\prime}$ are related by equation~(\ref{u'^2=u^2-I}).

Swapping in the symbolic argument of $W(\vec{V}\rightarrow\vec{V}^{\prime})$
the non-primed and primed variables,
\begin{linenomath*}
\begin{equation}
W(\vec{V}^{\prime}\rightarrow\vec{V})=mm_{\beta}\left(  m+m_{\beta}\right)
^{2}\left(  \frac{u^{\prime}}{u}\right)  G\left(  u^{\prime},\Lambda^{\prime
}\right)  , \label{W(P'->P)_via_d_sigma}%
\end{equation}
\end{linenomath*}
and applying $W(\vec{V}^{\prime}\rightarrow\vec{V})$ to the gain term,
$\hat{S}_{\mathrm{gain}}$, we obtain equation~(\ref{S_arr(W)}). Repeating the
same major steps as for $S_{\mathrm{dep}}$, we obtain
equation~(\ref{S_arr_traditional}).

The resultant kinetic equation with the inelastic collision operator
generalizes the standard kinetic equation with the elastic Boltzmann collision
operator \citep{Huang:Statistical87,Lifshitz:Physical1981}. Under
stationary conditions with neglected fields, this kinetic equation reduces to
equation~(\ref{D_t(f)}).

\section{Calculation of $J_{a}^{\infty}$ in the near
zone\label{Calculation of J in the near zone}}

In this appendix, we calculate the integral of equation~(\ref{J_a^b}) with
$b=\infty$ in the near zone, $a\ll\lambda_{T}^{(1)}$. The contribution of
$R^{\prime}\gtrsim\lambda_{T}^{(1)}$ in this integral is relatively small, so
that we can neglect all factors involving $R^{\prime}/\lambda_{T}^{(1)}$,
\begin{linenomath*}
\begin{equation}
\left.  J_{a}^{\infty}\right\vert _{a\ll\lambda_{T}^{(1)}}=2\int_{a}^{\infty
}\left(  1-\sqrt{1-\frac{r_{\mathrm{M}}^{2}}{\left(  R^{\prime}\right)  ^{2}}%
}\right)  \frac{R^{\prime}dR^{\prime}}{\sqrt{\left(  R^{\prime}\right)
^{2}-R_{0}^{2}}}. \label{reduced_to}%
\end{equation}
\end{linenomath*}
The lower integration limit must satisfy $a\geq\max(R_{0},r_{\mathrm{M}})$, so
that the convenient choice for the lower limit $a$ depends on whether the
entire straight-line ballistic trajectories of particles cross the meteoroid surface.

\subsection{Trajectories not crossing the meteoroid, $R_{0}\geq r_{\mathrm{M}%
}$}

For trajectories that do not cross the meteoroid surface, $R_{0}\geq
r_{\mathrm{M}}$, the integration variable $R^{\prime}$ satisfies $R^{\prime
}\geq a\geq R_{0}$. Introducing
\begin{linenomath*}
\begin{equation}
k=\frac{r_{\mathrm{M}}}{R_{0}}\leq1,\qquad y=\frac{R^{\prime}}{R_{0}}\geq
\frac{a}{R_{0}}\geq1, \label{k,xi}%
\end{equation}
\end{linenomath*}
we rewrite (\ref{reduced_to}) as
\begin{linenomath*}
\begin{equation}
\left.  J_{a}^{\infty}\right\vert _{R_{0}\geq r_{\mathrm{M}}}=2R_{0}\left.
I\right\vert _{R_{0}\geq r_{\mathrm{M}}},\qquad\left.  I\right\vert
_{R_{0}\geq r_{\mathrm{M}}}\equiv\int_{\frac{a}{R_{0}}}^{\infty}\frac
{y-\sqrt{y^{2}-k^{2}}}{\sqrt{y^{2}-1}}\ dy. \label{J_tempo}%
\end{equation}
\end{linenomath*}
The integral $\left.  I\right\vert _{R_{0}\geq r_{\mathrm{M}}}$ can be
expressed in terms of the incomplete elliptic integrals of the 1st and 2nd
kind \citep{Abramowitz:Handbook}. Since in the literature these integrals are
defined in several ways depending on the choice of the argument and parameter,
we will adhere to the following definitions and notations:
\begin{linenomath*}
\begin{equation}
\mathrm{F}\left(  x;k\right)  =\int_{0}^{x}\frac{dt}{\sqrt{\left(
1-t^{2}\right)  \left(  1-k^{2}t^{2}\right)  }},\qquad\mathrm{E}\left(
x;k\right)  =\int_{0}^{x}\sqrt{\frac{1-k^{2}t^{2}}{1-t^{2}}}\ dt, \label{F,E}%
\end{equation}
\end{linenomath*}
To express $\left.  I\right\vert _{R_{0}\geq r_{\mathrm{M}}}$ in terms of
$\mathrm{F}\left(  x;k\right)  $ and $\mathrm{E}\left(  x;k\right)  $ with
real $x$ and $k$ obeying $0\leq x,k<1$, we start by presenting the
corresponding indefinite integral as
\begin{linenomath*}
\begin{equation}
\int\frac{y-\sqrt{y^{2}-k^{2}}}{\sqrt{y^{2}-1}}\ dy=\sqrt{y^{2}-1}-\int%
\sqrt{\frac{y^{2}-k^{2}}{y^{2}-1}}\ dy. \label{step_1}%
\end{equation}
\end{linenomath*}
Temporarily changing in the second integral the variable $y$ to $x=1/y$, we
obtain
\begin{linenomath*}
\begin{align}
&  -\int\sqrt{\frac{y^{2}-k^{2}}{y^{2}-1}}\ dy=\int\frac{\sqrt{1-k^{2}x^{2}}%
}{x^{2}\sqrt{1-x^{2}}}\ dx\nonumber\\
&  =\int\frac{dx}{x^{2}\sqrt{\left(  1-x^{2}\right)  \left(  1-k^{2}%
x^{2}\right)  }}-k^{2}\int\frac{dx}{\sqrt{\left(  1-x^{2}\right)  \left(
1-k^{2}x^{2}\right)  }}. \label{step_2}%
\end{align}
\end{linenomath*}
Expressing the integrand of the first term as
\begin{linenomath*}
\begin{align}
&  \frac{1}{x^{2}\sqrt{\left(  1-x^{2}\right)  \left(  1-k^{2}x^{2}\right)  }%
}\nonumber\\
&  =\frac{k^{2}x^{2}}{\sqrt{\left(  1-x^{2}\right)  \left(  1-k^{2}%
x^{2}\right)  }}-\frac{d}{dx}\left[  \frac{\sqrt{\left(  1-x^{2}\right)
\left(  1-k^{2}x^{2}\right)  }}{x}\right]  , \label{via_derivative}%
\end{align}
\end{linenomath*}
we obtain
\begin{linenomath*}
\begin{align}
&  \int\frac{dx}{x^{2}\sqrt{\left(  1-x^{2}\right)  \left(  1-k^{2}%
x^{2}\right)  }}=\int\frac{dx}{\sqrt{\left(  1-x^{2}\right)  \left(
1-k^{2}x^{2}\right)  }}\nonumber\\
&  -\int\sqrt{\frac{1-k^{2}x^{2}}{1-x^{2}}}\ dx-\frac{\sqrt{\left(
1-x^{2}\right)  \left(  1-k^{2}x^{2}\right)  }}{x}. \label{step_3}%
\end{align}
\end{linenomath*}
Combining (\ref{step_1}), (\ref{step_2}), (\ref{step_3}) and returning to the
original variables, we obtain
\begin{linenomath*}
\begin{align}
&  \left.  J_{a}^{\infty}\right\vert _{R_{0}\geq r_{\mathrm{M}}}=2R_{0}\left.
I\right\vert _{R_{0}\geq r_{\mathrm{M}}}=2R_{0}\left[  \mathrm{E}\left(
\frac{R_{0}}{a};\frac{r_{\mathrm{M}}}{R_{0}}\right)  -\left(  1-\frac
{r_{\mathrm{M}}^{2}}{R_{0}^{2}}\right)  \mathrm{F}\left(  \frac{R_{0}}%
{a};\frac{r_{\mathrm{M}}}{R_{0}}\right)  \right] \nonumber\\
&  -2\left(  1-\sqrt{1-\frac{r_{\mathrm{M}}^{2}}{a^{2}}}\right)  \sqrt
{a^{2}-R_{0}^{2}}. \label{integ_prom}%
\end{align}
\end{linenomath*}

\subsection{Trajectories crossing the meteoroid, $R_{0}\leq r_{\mathrm{M}}$}

For trajectories that cross the meteoroid surface, $R_{0}\leq r_{\mathrm{M}}$,
we have $r_{\mathrm{M}}\leq a\leq R^{\prime}$. In this case we keep the same
temporary notations $k$ and $y$ as in (\ref{k,xi}) and (\ref{J_tempo}) but
with different definitions:
\begin{linenomath*}
\begin{equation}
k=\frac{R_{0}}{r_{\mathrm{M}}}\leq1,\qquad y=\frac{R^{\prime}}{r_{\mathrm{M}}%
}\geq\frac{a}{r_{\mathrm{M}}}\geq1. \label{k,z}%
\end{equation}
\end{linenomath*}
Then instead of (\ref{J_tempo}) we introduce
\begin{linenomath*}
\begin{equation}
\left.  J_{a}^{\infty}\right\vert _{R_{0}\leq r_{\mathrm{M}}}=2r_{\mathrm{M}%
}\left.  I\right\vert _{R_{0}\leq r_{\mathrm{M}}},\qquad\left.  I\right\vert
_{R_{0}\leq r_{\mathrm{M}}}\equiv\int_{\frac{a}{r_{\mathrm{M}}}}^{\infty}%
\frac{y-\sqrt{y^{2}-1}}{\sqrt{y^{2}-k^{2}}}\ dy. \label{J_tempo_2}%
\end{equation}
\end{linenomath*}
Following the same steps as for $R_{0}\geq r_{\mathrm{M}}$, we have:
\begin{linenomath*}
\begin{equation}
\int\frac{y-\sqrt{y^{2}-1}}{\sqrt{y^{2}-k^{2}}}\ dy=\sqrt{y^{2}-k^{2}}%
-\int\sqrt{\frac{y^{2}-1}{y^{2}-k^{2}}}\ dy, \label{step_4}%
\end{equation}
\end{linenomath*}
where the second term in the RHS we recast as
\begin{linenomath*}
\begin{align}
&  -\int\sqrt{\frac{y^{2}-1}{y^{2}-k^{2}}}\ dy\ ~\overset{y=\frac{1}%
{x}}{\overbrace{=}\ }\int\frac{\left(  1-x^{2}\right)  dx}{x^{2}\sqrt{\left(
1-x^{2}\right)  \left(  1-k^{2}x^{2}\right)  }}\nonumber\\
&  =\int\frac{dx}{x^{2}\sqrt{\left(  1-x^{2}\right)  \left(  1-k^{2}%
x^{2}\right)  }}-\int\frac{dx}{\sqrt{\left(  1-x^{2}\right)  \left(
1-k^{2}x^{2}\right)  }}. \label{prolka}%
\end{align}
\end{linenomath*}
For the first term in the RHS of (\ref{prolka}) we can use
equation~(\ref{via_derivative}). This yields
\begin{linenomath*}
\begin{equation}
-\int\sqrt{\frac{y^{2}-1}{y^{2}-k^{2}}}\ dy=-\ \frac{\sqrt{\left(
y^{2}-1\right)  \left(  y^{2}-k^{2}\right)  }}{y}-\left.  \int\sqrt
{\frac{1-k^{2}x^{2}}{1-x^{2}}}\ dx\right\vert _{x=\frac{1}{y}} \label{step_5}%
\end{equation}
\end{linenomath*}
Combining (\ref{step_4}) with (\ref{step_5}), we obtain
\begin{linenomath*}
\[
\int\frac{y-\sqrt{y^{2}-1}}{\sqrt{y^{2}-k^{2}}}\ dy=\left(  1-\sqrt{1-\frac
{1}{y^{2}}}\right)  \sqrt{y^{2}-k^{2}}-\left.  \int\sqrt{\frac{1-k^{2}x^{2}%
}{1-x^{2}}}\ dx\right\vert _{x=\frac{1}{y}}%
\]
\end{linenomath*}
so that
\begin{linenomath*}
\begin{equation}
\left.  J_{a}^{\infty}\right\vert _{R_{0}\leq r_{\mathrm{M}}}=2r_{\mathrm{M}%
}\left.  I\right\vert _{R_{0}\leq r_{\mathrm{M}}}=2r_{\mathrm{M}}%
\mathrm{E}\left(  \frac{r_{\mathrm{M}}}{a};\frac{R_{0}}{r_{\mathrm{M}}%
}\right)  -2\left(  1-\sqrt{1-\frac{r_{\mathrm{M}}^{2}}{a^{2}}}\right)
\sqrt{a^{2}-R_{0}^{2}}. \label{integ_prom_drugoj}%
\end{equation}
\end{linenomath*}


\end{document}